\documentclass[a4paper,11pt]{article}
\pdfoutput=1 

\usepackage[usenames,dvipsnames,svgnames,table]{xcolor} 
\usepackage{jcapmod}
\usepackage{verbatim}

\usepackage[T1]{fontenc} 

\definecolor{lightgray}{gray}{0.9}
\definecolor{Amber}{rgb}{1.0, 0.75, 0.0}
\definecolor{blizzardblue}{rgb}{0.67, 0.9, 0.93}

\usepackage[latin9]{inputenc}
\usepackage{float}
\usepackage{graphicx}
\usepackage{hyperref}
\usepackage{amsmath}
\usepackage{amssymb}
\usepackage{todonotes}
\usepackage[mathscr]{eucal}
\usepackage{microtype}
\usepackage{esint}
\usepackage{amsbsy}
\usepackage{color}
\usepackage{cleveref}
\usepackage{breakurl}
\usepackage[normalem]{ulem}
\usepackage{bbm}
\usepackage{mathtools}
\usepackage{braket}
\usepackage{comment}
\usepackage[super]{nth}
\usepackage{csquotes}
\usepackage{lmodern}

\numberwithin{equation}{section}
\usepackage{array}

\usepackage{siunitx}
\DeclareSIUnit \h {\mbox{$h$}}
\DeclareSIUnit \parsec {pc}

\usepackage{tikz}
\usepackage{pgfplots}
\pgfplotsset{compat=newest,every axis plot/.append style={line width=1pt}}

\makeatletter

\pdfpageheight\paperheight
\pdfpagewidth\paperwidth

\@ifundefined{textcolor}{}{%
 \definecolor{BLACK}{gray}{0}
 \definecolor{WHITE}{gray}{1}
 \definecolor{RED}{rgb}{1,0,0}
 \definecolor{GREEN}{rgb}{0,1,0}
 \definecolor{BLUE}{rgb}{0,0,1}
 \definecolor{CYAN}{cmyk}{1,0,0,0}
 \definecolor{MAGENTA}{cmyk}{0,1,0,0}
 \definecolor{YELLOW}{cmyk}{0,0,1,0}
 \definecolor{antiquefuchsia}{rgb}{0.57, 0.36, 0.51}
}

\newcommand{\shaderow}{\rowcolor{lightgray}}

\renewcommand\thesection{\arabic{section}}

\allowdisplaybreaks

\makeatother

\makeatletter

\DeclareRobustCommand{\rcite}[1]{%
  \rcite@aux#1,\@nil{#1}%
}
\def\rcite@aux#1,#2\@nil#3{%
  \if\relax#2\relax
    Ref.~~\cite{#3}%
  \else
    Refs.~~\cite{#3}%
  \fi
}
\makeatother

\title{Quintessential $\alpha$-attractor inflation: forecasts for Stage IV galaxy surveys}

\author[a,b]{\large Yashar Akrami,}
\author[c]{Santiago Casas,}
\author[a]{Senwen Deng,}
\author[d]{and Valeri Vardanyan}

\affiliation[a]{Laboratoire de Physique de l'\'Ecole Normale Sup\'erieure, ENS, Universit\'e PSL, CNRS, Sorbonne Universit\'e, Universit\'e de Paris, F-75005 Paris, France}
\affiliation[b]{Observatoire de Paris, Universit\'e PSL, Sorbonne Universit\'e, LERMA, 75014 Paris, France}
\affiliation[c]{AIM, CEA, CNRS, Universit\'e Paris-Saclay, Universit\'e Paris Diderot, Sorbonne Paris Cit\'e, F-91191 Gif-sur-Yvette, France}
\affiliation[d]{Kavli Institute for the Physics and Mathematics of the Universe (WPI), UTIAS, The University of Tokyo, Chiba 277-8583, Japan}

\emailAdd{akrami@ens.fr}
\emailAdd{santiago.casas@cea.fr}
\emailAdd{senwen.deng@ens.fr}
\emailAdd{valeri.vardanyan@ipmu.jp}

\abstract{Single-field models of $\alpha$-attractor quintessential inflation provide a unified picture of the two periods of early- and late-time cosmic acceleration, where both inflation and dark energy are described by a single scalar degree of freedom rolling down a runaway potential. These theoretically well-motivated models have distinct observational predictions that are in agreement with existing cosmological data. We show that the next generation of large-scale structure surveys, even when no other cosmological data sets are considered, will strongly constrain the parameter space of these models, and test them against the standard cosmological model and more conventional non-quintessential inflation. In particular, we expect $\mathcal{O}(10^{-5}\mathrm{-}10^{-4})$ constraints on the present values of the dark energy equation of state and its time derivative, $w_0$ and $w_a$. We also forecast more than one order of magnitude tighter constraints on the spectral index of primordial curvature perturbations $n_s$ compared to the expectations for the standard model. This demonstrates the powerful synergy between the upcoming large-scale structure probes of inflation and those aiming to measure the tensor-to-scalar ratio $r$ through the observation of $B$-mode polarization of the cosmic microwave background.}


\begin{document}
\maketitle

\section{Introduction}
 
Cosmology has become a science of synergies between conceptually different probes of the Universe. As the amount of cosmological data is growing exponentially and their quality is continuously improving across all wavelengths, these synergies have now become of increasing importance. Cosmological observations of completely different epochs in the cosmic history are now able to measure the very same physical quantities.

The measurement of the spectral index of primordial curvature perturbations, $n_s$, is an example of this synergistic approach. It has been measured primarily and with remarkable precision by the observations of the cosmic microwave background (CMB)~\cite{Akrami:2018odb,Aghanim:2018eyx}. However, as the spectral index is an essential characteristic of the initial conditions for dark matter clustering, upcoming high-precision observations of the cosmic large-scale structure (LSS) are expected to provide complementary constraints on $n_s$~\cite{Pourtsidou:2016ctq,Sprenger:2018tdb,Bacon:2018dui,Blanchard:2019oqi}.  

There is a similar synergistic approach at the level of cosmological model building. One particularly attractive avenue is constructing models that explain a wide span of cosmic epochs with limited sets of theoretical ingredients, some of which control both the early-Universe dynamics, such as cosmic inflation, and the late-time phenomena, such as the present-day cosmic acceleration. One compelling idea is the so-called {\it quintessential inflation}~\cite{Peebles:1998qn,Peloso:1999dm,Tashiro:2003qp,Majumdar:2001mm,Sahni:2001qp,Huey:2001ae,Dimopoulos:2001ix,Sami:2004xk,Hossain:2014xha}, where a single scalar field drives both of the accelerating phases of the Universe.

In this paper, we study {\it $\alpha$-attractor} models of quintessential inflation, a theoretically well-motivated class of models that have been proven to be able to sustain a successful inflationary phase and serve as interesting models of dark energy. $\alpha$-attractors~\cite{Kallosh:2013hoa,Kallosh:2013yoa,Galante:2014ifa,Kallosh:2014rga,Kallosh:2015lwa,Linde:2015uga,Carrasco:2015pla}, constructed primarily in the context of supergravity, have gained much attention in inflationary model building. Their power lies in their nearly universal and unifying predictions for properties of primordial fluctuations.

Motivated by their inflationary success, $\alpha$-attractors have also been considered as candidates for dark energy. Particularly, Ref.~\cite{Linder:2015qxa} studied dark energy models with poles in their kinetic terms, and demonstrated how a single model of this class could accommodate hybrid properties of {\it thawing} and {\it freezing} types. Interestingly, the pole structure also helps to reduce the level of fine tuning of initial conditions notoriously present in canonical scalar-field models of dark energy. Ref.~\cite{Bag:2017vjp} then investigated a broader class of models and classified their dynamical properties. Ref.~\cite{Garcia-Garcia:2018hlc} performed a Bayesian parameter estimation for the class of models proposed in Ref.~\cite{Linder:2015qxa}, while Ref.~\cite{Garcia-Garcia:2019ees} performed a Bayesian forecast analysis for the same class of models in view of upcoming galaxy surveys and CMB experiments. Ref.~\cite{Cedeno:2019cgr} performed Bayesian model comparison between the models of Ref.~\cite{Bag:2017vjp} and the $\Lambda$CDM model. Finally, dark energy with modified pole structure has been proposed and studied in Ref.~\cite{Linder:2019caj} with interesting speculations about their connection to the recently proposed swampland conjectures~\cite{Obied:2018sgi,Agrawal:2018own,Garg:2018reu,Ooguri:2018wrx}.\footnote{Note that the swampland conjectures are generically in strong tension with observationally viable single-field $\alpha$-attractor models, similarly to any other single-field models of slow-roll inflation or quintessence; see, e.g., Refs.~\cite{Akrami:2018ylq,Raveri:2018ddi,Achucarro:2018vey,Akrami:2020zfz}.}

All of these studies, however, have focused on models which, while being motivated by the inflationary framework of $\alpha$-attractors, are not necessarily connected to the inflationary epoch, i.e., they are not models of quintessential inflation. In the context of quintessential inflation, $\alpha$-attractor models have been proposed and studied in Refs.~\cite{Dimopoulos:2017zvq,Dimopoulos:2017tud,Akrami:2017cir}. In particular, Ref.~\cite{Akrami:2017cir} investigated a broad class of theoretically well-motivated scenarios, which included both single- and multi-field models. A generic feature of the single-field models of $\alpha$-attractor quintessential inflation, in contrast to more standard inflationary scenarios, is the appearance of monotonic potentials. Inflation takes place on the higher-energy parts of the potentials, while the low-energy tails support dark energy.

Ref.~\cite{Akrami:2017cir} has additionally demonstrated that these models provide novel universal predictions for inflationary parameters, which can serve as ``smoking gun'' signatures in distinguishing this class of models of inflation from the more conventional ones. As we detail in Section~\ref{sec:background}, one immediate implication of these quintessential inflationary models is the presence of a kinetic-dominated epoch after inflation, before the onset of standard radiation- and matter-dominated eras. This kinetic-dominated period results in a substantial increase of $n_s$, by $\sim0.006$, compared to that of inflationary models based on potentials with minima. Precise measurements of $n_s$ are therefore key in distinguishing quintessential models of inflation from non-quintessential ones. A significant reduction in the statistical error on the measured value of $n_s$ is expected in the next few years through the data provided by the next generation of cosmological surveys, observing both the CMB and the LSS~\cite{Finelli:2016cyd,Abazajian:2016yjj,Abazajian:2020dmr,Ade:2018sbj,Pourtsidou:2016ctq,Sprenger:2018tdb,Bacon:2018dui,Blanchard:2019oqi}.

Additionally, we stress that within quintessential $\alpha$-attractors several interesting interconnections exist between typically unrelated observables. Particularly, the primordial tensor-to-scalar ratio $r$ and the dark energy equation of state $w_\mathrm{DE}$ are both controlled by the parameter $\alpha$. Moreover, the spectral index $n_s$ and the amplitude of primordial scalar perturbations $A_s$ are determined by the parameters of the scalar field potential, hence they directly affect the late-Universe dynamics. These interconnections call for combining cosmological probes from different epochs of the Universe.\footnote{It should be mentioned that even though such interconnections are not common within dark energy and inflationary models, they are not unique to $\alpha$-attractors; see, e.g., Refs.~\cite{Rubio:2017gty,Geng:2017mic,Casas:2017wjh,Casas:2018fum,Haro:2018zdb,Camargo-Molina:2019faa,Dimopoulos:2019gpz,Dimopoulos:2019ogl,Benisty:2020qta} for other examples.}

The first observational analysis of models of $\alpha$-attractor quintessential inflation has been performed in Ref.~\cite{Akrami:2017cir}, where the viability of the models has been studied in light of the existing cosmological data and using geometrical constraints on the cosmic history at the background level. Motivated by the arguments above, we extend the analysis of Ref.~\cite{Akrami:2017cir} to the question of how well upcoming, next-generation cosmological surveys will constrain those classes of $\alpha$-attractor quintessential inflation.

In the next decade, several CMB and LSS surveys will provide a large amount of new data with unprecedented precision. These so-called Stage IV surveys include the {\it Euclid} space telescope~\cite{Laureijs:2011gra,Blanchard:2019oqi}, the Dark Energy Spectroscopic Instrument (DESI)~\cite{Aghamousa:2016zmz,Aghamousa:2016sne}, the Rubin Observatory Legacy Survey of Space and Time (LSST)~\cite{Ivezic:2008fe,Abell:2009aa,Mandelbaum:2018ouv}, the Square Kilometre Array (SKA)~\cite{Yahya:2014yva,Bacon:2018dui},
the Wide Field InfraRed Survey Telescope (WFIRST)~\cite{Spergel:2015sza}, the Subaru Hyper Suprime-Cam (HSC) and Prime Focus Spectrograph (PFS) surveys~\cite{Aihara:2017paw,Tamura:2016wsg} and the Spectro-Photometer for the History of the Universe, Epoch of Reionization, and Ices Explorer (SPHEREx)~\cite{Dore:2014cca,Dore:2018kgp}, all measuring the distribution and evolution of the large-scale structure, as well as LiteBIRD~\cite{Matsumura:2013aja,Hazumi:2019lys}, CMB-S4~\cite{Abazajian:2016yjj,Abazajian:2020dmr}, CMB-HD~\cite{Sehgal:2019ewc,Sehgal:2020yja} and the Simons Observatory (SO)~\cite{Ade:2018sbj}, observing the cosmic microwave background.

In this paper, we focus on forecasts for upcoming Stage IV LSS surveys, use expected measurements of {\it galaxy clustering} and {\it weak gravitational lensing} by next-generation galaxy surveys, and perform a Fisher forecast analysis for each one and combinations of the two probes to estimate the upcoming constraints on $\alpha$-attractor models of quintessential inflation. We present our results for two sets of LSS surveys, one of which is a combination of the DESI and LSST surveys and the other one is SKA2, the second phase of the SKA. We show that in both cases, the upcoming LSS surveys will strongly constrain the parameter space of the $\alpha$-attractor models. We demonstrate that the expected constraints on some derived parameters, such as the scalar spectral index $n_s$, are between one and two orders of magnitude stronger than the anticipated constraints in the framework of $\Lambda$CDM. We also show that the equation of state of dark energy will be tightly constrained by future observations. These all prove the power of the next-generation cosmological surveys in constraining one important class of inflationary and dark energy models and testing them against the standard $\Lambda$CDM model and conventional models of non-quintessential inflation.

The paper is organized as follows. In Section~\ref{sec:alphaattractors}, we first review briefly the theory of cosmological $\alpha$-attractors. We present in Section~\ref{sec:background} cosmological solutions at the level of the background, as well as some of their interesting implications for early and late times. In Section~\ref{sec:pert}, we provide equations for the evolution of cosmic structure at the linear level and the growth of cosmological perturbations. Section~\ref{sec:mod-params} presents the specific models of $\alpha$-attractor quintessential inflation that we study in this paper, as well as the parameter space of the models that we explore in our Fisher forecast analysis. In Section~\ref{sec:fisherformalism}, we first review basics of the Fisher matrix formalism, and in Sections~\ref{sec:GC} and~\ref{sec:WL} we discuss the two probes of galaxy clustering and weak lensing, respectively, and provide all the expressions needed for performing the Fisher forecasts. After describing the Stage IV galaxy surveys considered in this paper and their specifications in Section~\ref{sec:GS-spec}, we detail the fiducial parameters we assume for the studied $\alpha$-attractor models in Section~\ref{sec:fid-Fisher}. We then present the results of our forecast analysis, i.e., the predicted constraints on the parameters of the models, as well as on the derived cosmological parameters. We conclude in Section~\ref{sec:conclusions}.

\section{Single-field $\alpha$-attractor quintessential inflation}\label{sec:alphaattractors}

The essential, underlying ingredient for the universality of $\alpha$-attractor models is their non-canonical and singular kinetic term, rooted in hyperbolic field-space metric. Often extra directions in this hyperbolic field space are taken to be stabilized, and the $\alpha$-attractors are described as a single-field model with the Lagrangian density
\begin{equation}\label{eq:Lagrangian1}
    \mathcal{L}=\sqrt{-g}\left(\frac{1}{2}R - \frac{1}{2}\frac{\left(\partial_{\mu} \phi\right)^{2}}{\left(1-\frac{\phi^{2}}{6 \alpha}\right)^{2}}-V(\phi)\right) + \mathcal{L}_\mathrm{matter}\,,
 \end{equation}
where $g$ is the determinant of the space-time metric $g_{\mu\nu}$, $R$ is the corresponding Ricci scalar, and $\phi(x,t)$ is a scalar field with a potential $V(\phi)$.  $\mathcal{L}_\mathrm{matter}$ denotes the matter Lagrangian density, decoupled from the scalar sector and minimally coupled to gravity;  it is unimportant for the inflationary phase and affects the dynamics of the Universe only in later epochs. Note that we have set the Planck mass $M_{\mathrm{Pl}}$ to unity. Finally, $\alpha$ is a positive constant related to the curvature of field space, and is a quantity of central importance in $\alpha$-attractors.

Inflation takes place in the immediate vicinity of the metric's singularity, and as such, its predictions are rather insensitive to the global details of the scalar-field potential $V(\phi)$. In particular, for $\alpha\sim\mathcal{O}(1)$ and smaller, predictions for the scalar spectral index $n_{s}$ and the tensor-to-scalar ratio $r$ are universal and given by
\begin{equation}\label{eq:rn_s}
	n_{s} =  1 - {2\over N_*}\, \qquad \mathrm{and} \qquad r= {12\alpha\over N_*^{2}}\,, 
\end{equation}
with $N_*$ denoting the number of $e$-foldings between the horizon crossing of modes of interest and the end of inflation. These predictions have been shown to be of perfect agreement with existing observational data~\cite{Akrami:2018odb,Aghanim:2018eyx}.

The theory described by (\ref{eq:Lagrangian1}) can be more conveniently studied in terms of the canonically-normalized field variable $\varphi$ defined through
\begin{equation}\label{eq:canonical-transformation}
    \phi\equiv\sqrt{6 \alpha} \tanh \frac{\varphi}{\sqrt{6 \alpha}}\,.
\end{equation}

We note that for $\alpha \rightarrow \infty$ the original field $\phi$ reduces to a canonically normalized variable. For finite $\alpha$, the poles $\phi^2 = 6\alpha$ correspond to $\varphi \rightarrow \pm \infty$. Importantly, in terms of $\varphi$ the potential $V(\sqrt{6 \alpha} \tanh \frac{\varphi}{\sqrt{6 \alpha}})$ develops flat directions, which are able to sustain accelerated expansion. The energy scales of these plateaus are given by the values of the original potential at the poles,
\begin{equation}
V_{\pm} \equiv V(\phi)|_{\phi = \pm \sqrt {6 \alpha}} \,.
\end{equation}

In standard, non-quintessential inflationary $\alpha$-attractor models, $V(\phi)$ has a minimum, the canonical field $\varphi$ rolls down the potential on one of the two plateaus $V_{\pm}$ towards the minimum, and the Universe undergoes an inflationary phase. The field then reaches the minimum, starts oscillating, and reheats the Universe. The simplest example of this class of potentials is the quadratic $V(\phi)=m^2\phi^2/2$.

The potential may however be of a runaway form~\cite{Akrami:2017cir,Dimopoulos:2017zvq}, such as the linear and exponential potentials, $V(\phi)\propto\phi$ and $V(\phi)\propto e^{\gamma\phi}$. These potentials are of special interest as they can describe both the early-time inflationary period and the late-time cosmic acceleration, meaning that the same scalar field plays the roles of both inflaton and quintessence---the models are therefore referred to as models of {\it $\alpha$-attractor quintessential inflation}.

As far as the value of $\alpha$ is concerned, it can be considered as a free phenomenological parameter of $\alpha$-attractor models, to be determined observationally. There are however certain values for $\alpha$ which are theoretically of particular interest. In the context of supersymmetry and supergravity, $\alpha$-attractors can be obtained by consistently reducing $\mathcal{N}=8$ supersymmetry (e.g., M-theory in $D=11$, string theory in $D=10$, or maximal supergravity in $D=4$) to minimal $\mathcal{N}=1$ supersymmetry. In this case, $\alpha$ takes the specific values~\cite{Ferrara:2016fwe,Kallosh:2017ced,Kallosh:2017wnt}
\begin{equation}
    3 \alpha\in\{1,2,3,4,5,6,7\}\,,\label{eq:alpha-range}
\end{equation}
while its value is arbitrary in constructions based directly on $\mathcal{N}=1$ supersymmetry.

\subsection{Background dynamics}\label{sec:background}

Assuming, as usual, a flat Friedmann-Lema\^itre-Robertson-Walker (FLRW) metric, the background dynamics of the system is described by 
\begin{align}
    H^{2}\left( 3 - \frac{1}{2} \varphi^{\prime 2}\right) = V(\varphi)+\bar\rho_{\mathrm{M}}+\bar\rho_{\mathrm{R}}\,,\label{eq:Hsquare}\\
    {\varphi^{\prime \prime} + (3 - \epsilon) \varphi^{\prime} + \frac{1}{H^{2}} \frac{\partial V(\varphi)}{\partial \varphi} =0}\,,\label{eq:phiEqn} 
\end{align}
where $\bar\rho_\text{M}$ and $\bar\rho_\text{R}$ are the background energy densities of matter and radiation, respectively, and a prime denotes a derivative with respect to the number of $e$-foldings $N\equiv\ln a$, with $a$ the scale factor of the Universe. $H\equiv\dot a/a$ is the Hubble expansion rate, with an overdot denoting a derivative with respect to cosmic time. The Hubble slow-roll parameter $\epsilon$ is given by
\begin{equation}\label{eq:epsilonDef}
    \epsilon \equiv -\frac{H^{\prime}}{H} = \frac{1}{2}\left(\varphi^{\prime 2}-\frac{\bar\rho_{\mathrm{M}}^{\prime}+\bar\rho_{\mathrm{R}}^{\prime}}{3 H^{2}}\right)\,.
\end{equation}

The densities $\bar\rho_\text{M}$ and $\bar\rho_\text{R}$ should be neglected in Eqs.~(\ref{eq:Hsquare}) and (\ref{eq:phiEqn}) in the inflationary phase, while they are given by
\begin{align}
    \bar\rho_\mathrm{M} = 3H_0^2\Omega_\mathrm{M}^0e^{-3N}\, \qquad \mathrm{and} \qquad \bar\rho_\mathrm{R} = 3H_0^2\Omega_\mathrm{R}^0e^{-4N}\,,
\end{align}
at late times. Here, $H_0$ is the present value of the Hubble expansion rate, and $\Omega_\mathrm{M}^0$ and $\Omega_\mathrm{R}^0$ are, respectively, the matter and radiation density parameters today (i.e., at $N=0$).

For quintessential inflationary models, which are the subject of the studies here, the scalar field $\varphi$ describes both the inflationary and dark energy phenomena. Assuming a runaway potential for the field $\varphi$, it slowly rolls down the potential along the higher plateau $V_+$, resulting in a slow-roll inflationary epoch, towards the lower plateau $V_-$, accounting for dark energy at late times. The transition from early to late times is however of particular importance for this class of models, as it is significantly different from that of the standard inflationary scenarios. As stated above, since the potential does not possess a local minimum in quintessential inflation, the field cannot oscillate around a minimum and the standard picture of reheating after inflation is no longer valid. However, there are several other reheating mechanisms for quintessential inflationary models which make it possible for the energy of the field to be transferred to standard particles. These mechanisms have been discussed in detail in Refs.~\cite{Akrami:2017cir,Dimopoulos:2017zvq,Dimopoulos:2017tud}.

It can be shown that in these classes of quintessential inflation, and independently of the reheating mechanism, the scalar field goes through a so-called {\it kination} phase at the end of inflation, where the kinetic energy of the field dominates its potential energy. This gives rise to an equation of state $w=+1$ instead of the usual $w \approx 0$ for non-quintessential inflaton oscillating at the bottom of its potential.

Let us emphasize here a distinct characteristic of quintessential inflation compared to non-quintessential scenarios, which is a direct consequence of the kination phase. It is well known that the quantity $N_*$ is determined by~\cite{Planck:2013jfk,Akrami:2018odb}
\begin{equation}\label{eq:terminN}
N_\ast \simeq67 - \ln\left(\frac{k_*}{a_0H_0}\right) +\frac{1}{4}\ln\left(\frac{V_*^2}{\rho_{\rm end}}\right)-\frac{1}{12}\ln(g_\mathrm{reh})+\frac{1-3w_\mathrm{int}}{12(1+w_\mathrm{int})}\ln{\left(\frac{\rho_{\rm reh}}{\rho_{\rm end}} \right)}\,,
\end{equation}
where $k_*=0.002~\mathrm{Mpc}^{-1}$ is the pivot scale, $a_0H_0$ is the present Hubble scale, $V_*$ is the potential energy when $k_*$ crossed the Horizon during inflation, $\rho_{\rm end}$ is the energy density at the end of inflation, $g_\mathrm{reh}$ is the number of effective bosonic degrees of freedom at the reheating energy scale $\rho_{\rm reh}$, and $w_\mathrm{int}$ is the effective equation of state during reheating. Since in quintessential inflation the inflaton/quintessence field goes through a kination phase during reheating with $w_\mathrm{int}\approx+1$ (as opposed to $w_\mathrm{int}\approx0$ for non-quintessential inflation), the last term in Eq.~(\ref{eq:terminN}) results in a difference of $\sim10$ in $N_*$ compared to conventional inflationary models. This consequently leads to an increase of $\sim0.006$ in the spectral index $n_s$, as follows from Eqs.~(\ref{eq:rn_s})~\cite{Akrami:2017cir}. Standard, non-quintessential inflationary potentials typically require an $N_*\sim50$ (i.e., an $n_s\sim0.960$), while $N_*\sim60$ (i.e., $n_s\sim0.966$) for quintessential potentials. The exact values of $N_*$, and therefore $n_s$, depend on the details of the model.

Kinetic energy density is diluted faster than radiation density, and the Universe eventually transfers from a kination-dominated epoch to a radiation-dominated one. It can be shown~\cite{Dimopoulos:2017zvq,Akrami:2017cir} that the scalar field $\varphi$ moves a maximum distance of $\sim43M_\mathrm{Pl}$ after the end of inflation and then freezes due to Hubble friction. This maximum excursion corresponds to {\it gravitational reheating}, which is the least efficient reheating mechanism. Clearly, a more efficient reheating, such as {\it instant preheating}, reduces the scalar field's excursion before freezing. After reheating, the Universe undergoes a radiation domination period followed by matter domination. The field remains frozen at a certain value $\varphi_\text{F}$ for this entire period after reheating until the Hubble friction reduces significantly, when the field starts rolling down its potential again. This is the onset of the dark energy domination epoch.

In summary, the background evolution of the Universe determined by quintessential $\alpha$-attractor inflation can be presented schematically by the sequence
\begin{align}\label{eq:sequence}
\text{inflation}\rightarrow\text{kination}\rightarrow\text{radiation}\rightarrow\text{dark matter}\rightarrow\text{dark energy}\,.
\end{align}

\subsection{Cosmological perturbations and growth of structure}\label{sec:pert}

The focus of our studies in this paper is on the constraints that the upcoming, next-generation galaxy surveys are expected to impose on $\alpha$-attractor quintessential inflation. This means that we need to know how the cosmological large-scale structure evolves and grows in these models. Focusing on the linear regime, this means that we need to know the dynamics of the linear cosmological perturbations.

Similarly to any other canonically normalized and minimally coupled scalar field theory, the sound speed of the $\alpha$-attractor field $\varphi$ equals unity in its rest frame (see, e.g., Refs.~\cite{Ballesteros:2010ks,amendola_dark_2010}), and consequently, pressure suppresses the growth of scalar-field over-densities at scales smaller than the cosmological horizon~\cite{Ballesteros:2010ks,Tsujikawa:2013fta,amendola_dark_2010}. This means that the scalar-field perturbations are negligible compared to matter perturbations.

As in standard cosmology, Fourier modes $k$ of matter density contrast, $\delta_\text{M}\equiv\left(\rho_\text{M}-\bar\rho_\text{M}\right)/\bar\rho_\text{M}$, evolve according to
\begin{equation}\label{eq:delta}
\ddot\delta_{\text{M},k}+2H\dot\delta_{\text{M},k}-\frac{3}{2}\frac{H_0^2\Omega_\mathrm{M}^0}{a^3}\delta_{\text{M},k}=0\,.
\end{equation}
Defining the growth factor $D(t)$ through $\delta_\mathrm{M}(x, t)=D(t) \delta_\mathrm{M}(x, 0)$, with $\delta_\mathrm{M}(x, 0)$ the initial value of $\delta_\mathrm{M}$, $D$ satisfies the same equation of motion as $\delta_\mathrm{M}$. Rewriting Eq.~(\ref{eq:delta}) in terms of the growth rate $f$, defined as 
\begin{equation}
        f(a)\equiv\frac{\mathrm{d} \ln D(a)}{\mathrm{d} \ln a}\label{eqn:def_f}\,,
\end{equation}
one obtains
\begin{equation}
        f^\prime+f^{2}+\left(2+\frac{H^\prime}{H}\right) f=\frac{3}{2} \Omega_{\mathrm{M}}\label{eqn:f}  \, ,
\end{equation}
where $\Omega_\mathrm{M} = \Omega_\mathrm{M}^0 (H/H_0)^{-2}  a^{-3}$. Note that as we neglect scalar-field perturbations, an $\alpha$-attractor model follows the same growth equation as in the $\Lambda$CDM model, even though the evolution of the Hubble rate $H$ may be significantly different from that of $\Lambda$CDM---this then leads in general to different growth rates for the two models.

\subsection{Models and parameters}\label{sec:mod-params}

In order to specify a single-field model of quintessential inflation, one needs to choose a runaway potential for the scalar field. Several possible choices have been presented and studied in detail in Ref.~\cite{Akrami:2017cir}. Remarkably, a simple linear potential $V(\phi) = \gamma\phi + \Lambda$ can be successfully employed for our purposes. As discussed in Ref.~\cite{Akrami:2017cir}, dark energy phenomenology in this model is very close to that of the $\Lambda$CDM model. For that reason, and in order to allow for the possibility of alternative cosmological backgrounds in our analysis, we choose to restrict ourselves, without loss of generality, to exponential potentials of the form~\cite{Dimopoulos:2017zvq,Akrami:2017cir}
\begin{equation}
    V(\phi)=M^{2} e^{\gamma\left(\frac{\phi}{\sqrt{6 \alpha}}-1\right)}+V_{0}\,,\label{eq:Vphi}
    \end{equation}
where $M^2$, $\gamma$ and $V_0$ are free parameters. This is an important and representative class of potentials for quintessential inflation, and we refer the reader to Ref.~\cite{Akrami:2017cir} for a detailed discussion of their properties.

\begin{figure}
    \centering
    \includegraphics[width=0.65\columnwidth]{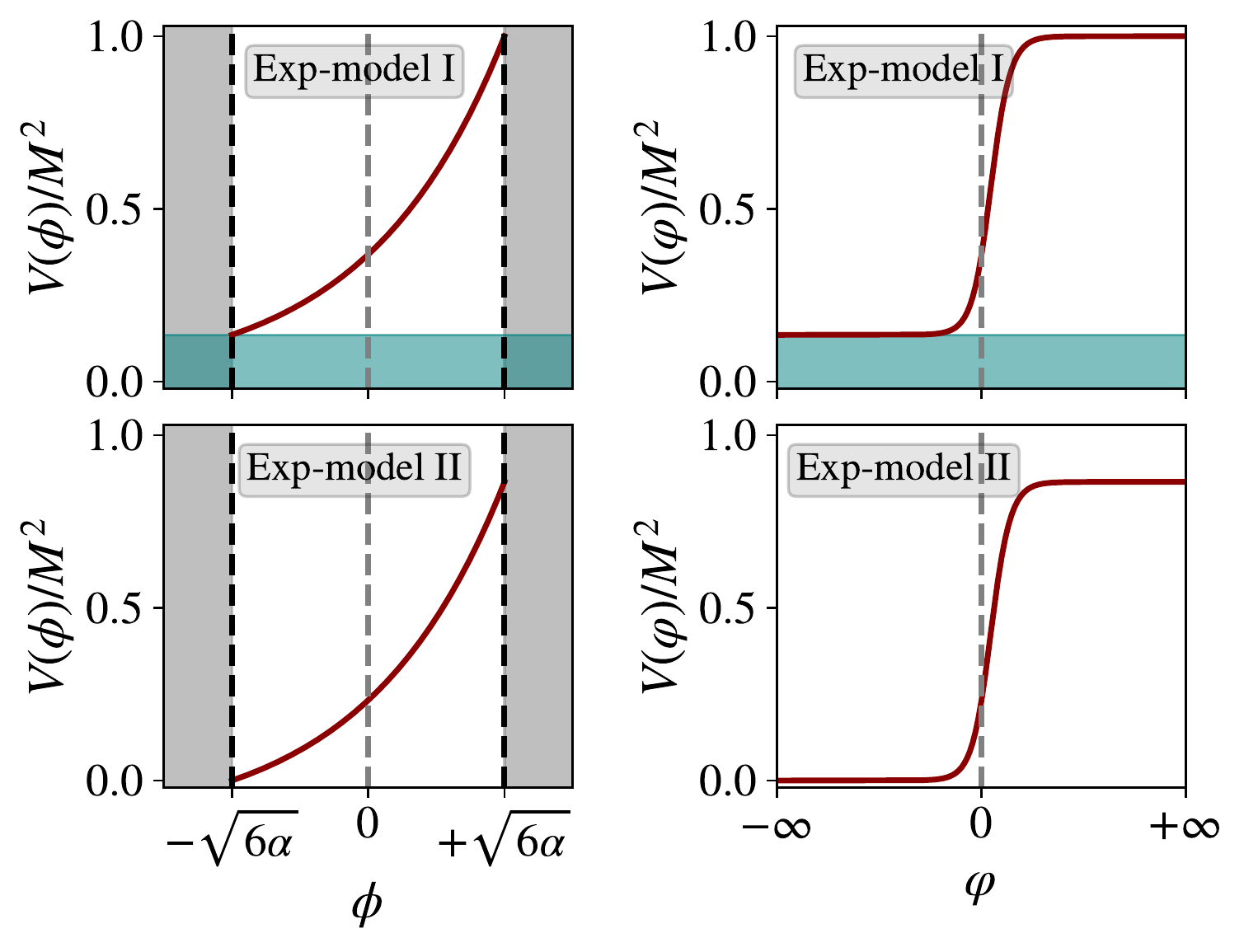}
    \caption{Schematic view of potentials for the two models of $\alpha$-attractor quintessential inflation considered in this paper:  {\it Exp-model I} (upper panels) and {\it Exp-model II} (lower panels). The left panels show the potentials in terms of the original scalar field $\phi$ (confined in the interval $(-\sqrt{6\alpha},\sqrt{6\alpha})$ between the two grey, shaded regions), while the right panels depict them in terms of the canonically normalized scalar field $\varphi$. {\it Exp-model I} has an asymptotic non-zero cosmological constant $\Lambda=M^{2} e^{-2 \gamma}$ at large negative $\varphi$ (shown by the green, shaded region), while the asymptotic cosmological constant vanishes for {\it Exp-model II}. Note that we have used $\gamma=1$ for the potentials here, which is different from the values of $\sim125$ in the actual cases studied in this paper; the depicted potentials are for illustration purposes only.} \label{fig:potentials}
\end{figure}

In terms of the canonically normalized scalar field $\varphi$, the potential is expressed as
\begin{equation}
    V(\varphi)=M^{2} e^{\gamma\left(\tanh \frac{\varphi}{\sqrt{6 \alpha}}-1\right)}+V_{0}\,,\label{eq:Vvarphi}
\end{equation}
with an inflationary plateau $V_{+} = M^2 +V_{0}$ for large positive $\varphi$ and a dark energy plateau with an asymptotic cosmological constant $\Lambda =V_{-}=M^2 e^{-2\gamma}+V_{0}$ for large negative $\varphi$. We focus in this paper on two specific cases with $V_0=0$ and $V_0=-M^{2} e^{-2 \gamma}$, which we call {\it Exp-model I} and {\it Exp-model II}, respectively, following the conventions of Ref.~\cite{Akrami:2017cir}. As illustrated in Fig.~\ref{fig:potentials}, the former possesses an asymptotic non-zero cosmological constant $\Lambda=M^{2} e^{-2 \gamma}$, while the asymptotic $\Lambda$ vanishes for the latter.

As has been discussed in Ref.~\cite{Akrami:2017cir}, in these models the parameter $M$, which determines the scale of inflation, is related to the amplitude of the power spectrum of primordial scalar perturbations $\mathcal{A}_s$ and the number of $e$-foldings before the end of inflation $N_*$ through the so-called COBE/Planck normalization (see, e.g., Ref.~\cite{Lyth:1998xn}) equation
\begin{equation}\label{eq:cobe}
    M^{2}=\frac{144 \pi^{2} \alpha N_*}{(2 N_*-3 \alpha)^{3}} \mathcal{A}_s\,.
    \end{equation}
$N_*$ is, on the other hand, related to the primordial scalar tilt $n_s$ through Eq.~(\ref{eq:rn_s}). This means that by knowing the values of $\alpha$, $n_s$ and $\mathcal{A}_s$, we can fully determine the value of the parameter $M$. Since $\mathcal{A}_s$ and $n_s$ are two of the six parameters of the standard concordance model, and in order to simplify comparison of our results with that of $\Lambda$CDM, we include $\mathcal{A}_s$ and $n_s$ (instead of $M$ and $N_*$) in the set of primary free parameters of $\alpha$-attractor models to be constrained by data.

In order to determine the dynamics of the scalar field in the late Universe, we should specify its initial conditions. We can always set the initial velocity to zero, as the Hubble friction before dark energy domination would suppress the velocity, even if it were set to a non-zero value initially. The choice of the initial value is somewhat more subtle. As we mentioned earlier, the field freezes after the reheating period, and this freezing value $\varphi_\mathrm{F}$ depends on the details of the reheating mechanism. As shown in Ref.~\cite{Akrami:2017cir}, depending on the value of $\varphi_\mathrm{F}$, the present equation of state of dark energy may deviate significantly from that of the $\Lambda$CDM model. In general, the smaller the $|\varphi_\mathrm{F}|$, the larger the deviation from $w=-1$. Since the exact value of $\varphi_\mathrm{F}$ is the subject of detailed analysis of the reheating epoch, for the purposes of this paper we assume it is known.

\begin{figure}
    \centering
    \includegraphics[width=0.6\columnwidth]{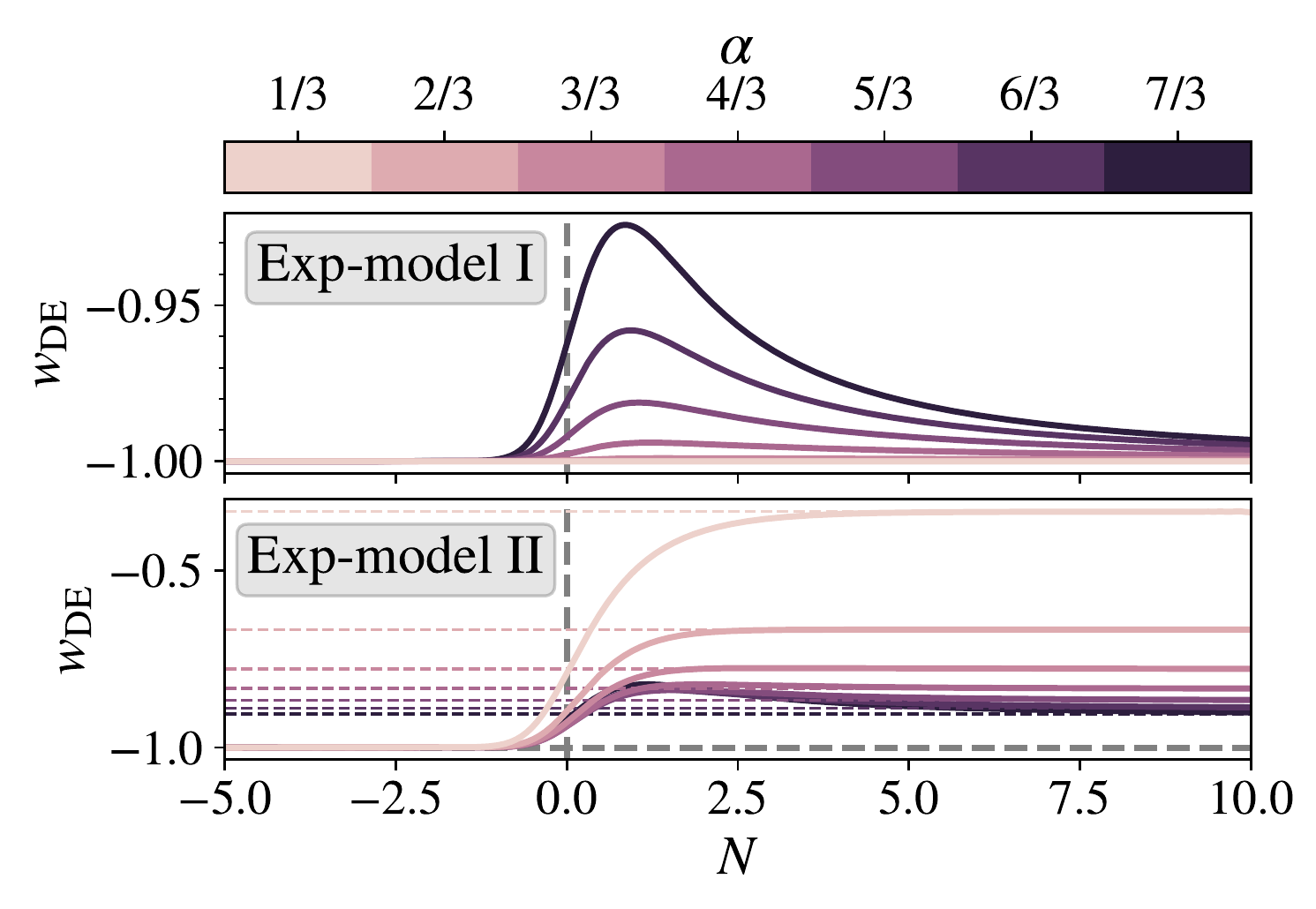}
    \caption{Time evolution of the dark energy equation of state $w_\mathrm{DE}$ for the two $\alpha$-attractor models of quintessential inflation, Exp-model I and Exp-model II, and for a range of theoretically-motivated values of $\alpha$. Time is measured in terms of the number of $e$-foldings $N \equiv \ln(a)$, with $N=0$ corresponding to the present. The parameters used to generate the figure are the fiducial parameters described later in the paper (see Table~\ref{table:fidu}). The initial velocity of the scalar field is set to $0$, while the initial value is chosen to be $-10M_\mathrm{Pl}$. The colored dashed horizontal lines in the lower panel correspond to the asymptotic values given by the second equation in~(\ref{eq:winf}).} \label{fig:wDE}
\end{figure}

The two specific models Exp-model I and Exp-model II described above represent, respectively, two generic classes of quintessential inflationary models with either
\begin{equation}\label{eq:winf}
w_{\infty}= -1\, \qquad \mathrm{or} \qquad w_{\infty}= -1+  \frac{2}{9\alpha}
\end{equation}
for the asymptotic value of the dark energy equation of state $w_\mathrm{DE}$. Fig.~\ref{fig:wDE} shows\footnote{This and several other figures in this paper can be reproduced using the code available at \url{https://github.com/valerivardanyan/alpha-attractor-DE}. The codes used for the main statistical analysis are described in Section~\ref{sec:fid-Fisher}.} examples of time evolution of $w_\mathrm{DE}$ for the two $\alpha$-attractor models Exp-model I and Exp-model II, and for a range of values of $\alpha$ given by~(\ref{eq:alpha-range}). Both panels demonstrate that $w_\mathrm{DE}\approx-1$ during the radiation- and matter-dominated epochs, where the scalar-field is frozen due to a large Hubble friction. As expected, it is only at late times that $w_\mathrm{DE}$ starts deviating from $-1$, resulting in values of $w_0$ and $w_a$, the two Chevallier-Polarski-Linder (CPL) parameters~\cite{Chevallier:2000qy,Linder:2002et}, different from those of the $\Lambda$CDM values $-1$ and $0$, respectively. The figure also shows that the asymptotic values of $w_\mathrm{DE}$ follow Eqs.~(\ref{eq:winf}). For Exp-model I, larger values of $\alpha$ give rise to larger deviations from $\Lambda$CDM at all times after the onset of dark energy domination, in terms of both $w_0$ and $w_a$. In this model, the scalar field starts rolling when dark energy becomes the dominant component, before asymptotically reaching the constant plateau. The case for Exp-model II is similar for large values of $\alpha$. This is expected because the asymptotic value of $w_\mathrm{DE}$ is close to $-1$ for these values; see the second equation in~(\ref{eq:winf}). For smaller values of $\alpha$, on the other hand, the asymptotic value of $w_\mathrm{DE}$ is so different from $-1$ that the field cannot overshoot it at the present time. As a result, $w_0$ starts increasing as $\alpha$ decreases. 

\begin{figure}
    \centering
    \includegraphics[width=0.6\columnwidth]{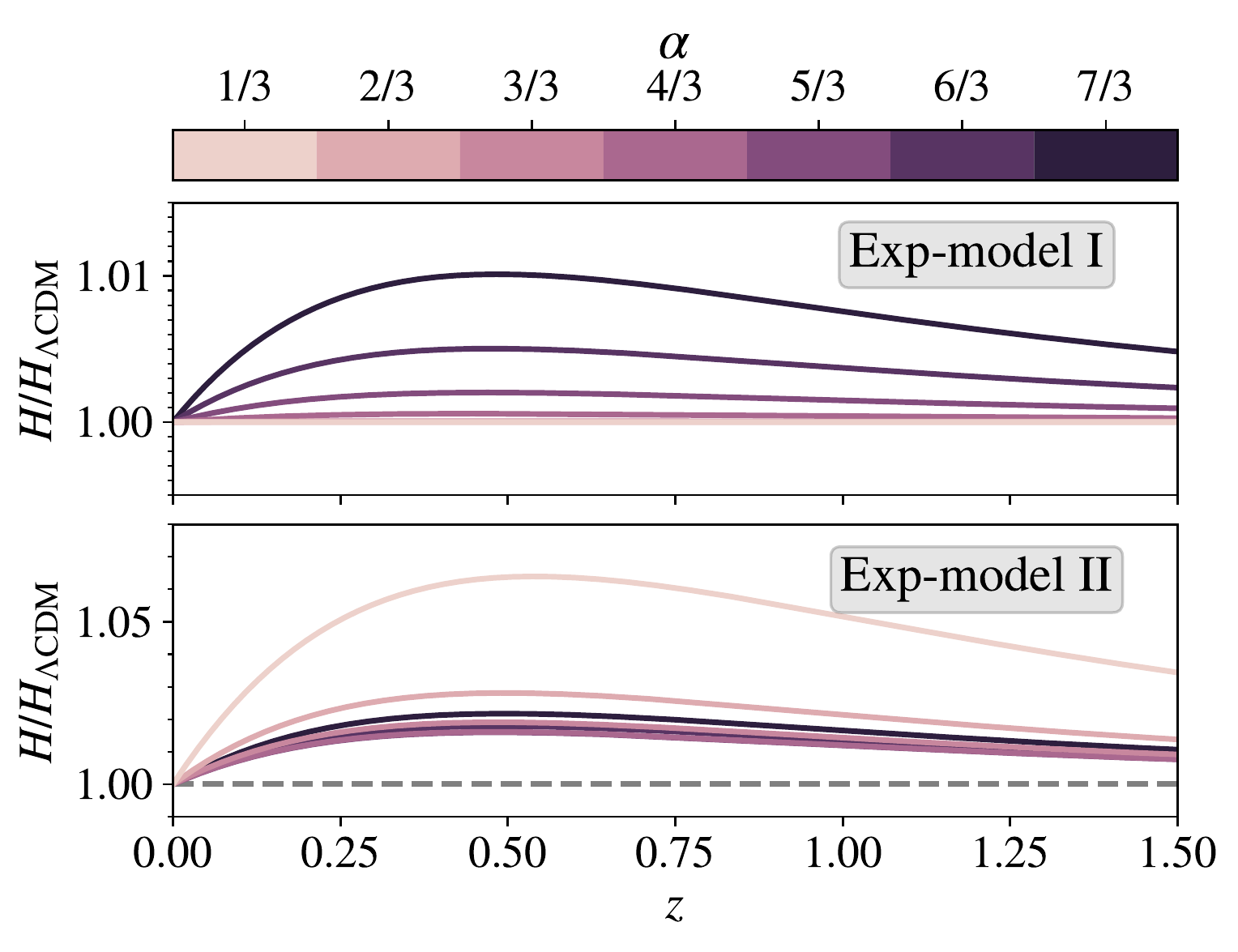}
    \caption{Redshift dependence of the Hubble expansion rate $H$ for the two $\alpha$-attractor models of quintessential inflation, Exp-model I and Exp-model II, compared to that of the $\Lambda$CDM model. The choice of parameters is the same as in Fig.~\ref{fig:wDE}.}\label{fig:H}
\end{figure}

\begin{figure}
    \centering
    \includegraphics[width=0.6\columnwidth]{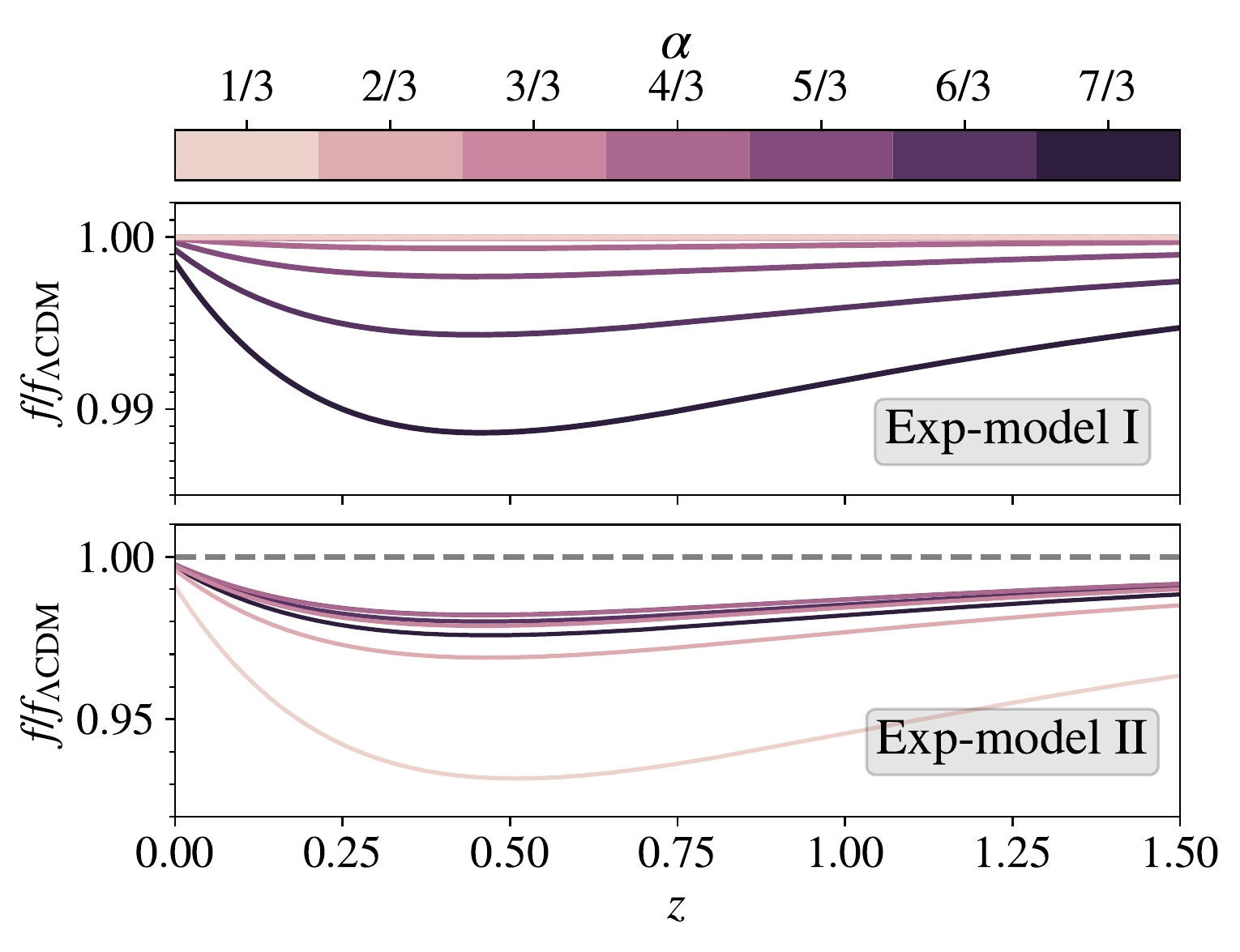}
    \caption{Redshift dependence of the growth rate $f$ for the two $\alpha$-attractor models of quintessential inflation, Exp-model I and Exp-model II, compared to that of the $\Lambda$CDM model. The choice of parameters is the same as in Fig.~\ref{fig:wDE}.}\label{fig:f}
\end{figure}

In Figs.~\ref{fig:H} and~\ref{fig:f}, we show the time evolution of the Hubble expansion rate $H$ and the growth rate $f$ of the cosmological perturbations for the same two models, in comparison to those of the $\Lambda$CDM model. The figures show that values of $H$ and $f$ are, respectively, larger and smaller for both $\alpha$-attractor models compared to those of $\Lambda$CDM over the entire range of redshifts probed by LSS surveys and for all values of $\alpha\in\{1/3,2/3,1,4/3,5/3,2,7/3\}$. We can also see that, similarly to the case of dark energy equation of state $w_\mathrm{DE}$, while the differences between the values of $H$ and $f$ for Exp-model I and those of the $\Lambda$CDM model both increase by increasing $\alpha$, we first see a decrease and then a slight increase in both $H$ and $f$ for Exp-model II.

It should be noted that we have enforced the models to give $h = 0.72$, where $h$ is the reduced Hubble expansion rate today defined through $H_0=\SI{100}{\h\kilo\meter\per\second\per\mega\parsec}$. 

\section{Fisher forecasts for next-generation galaxy surveys}\label{sec:fisher}

\subsection{Fisher formalism}\label{sec:fisherformalism}

Our goal in this paper is to estimate how well the upcoming Stage IV surveys of the large-scale structure will constrain $\alpha$-attractor models of quintessential inflation. More concretely, we aim to forecast the $68.3\%$ and $95\%$ confidence regions and intervals provided by the next generation of galaxy surveys around some fiducial model parameters. We do this by performing a {\it Fisher matrix} analysis.

For a given model, the Fisher formalism provides a computationally inexpensive way to forecast confidence regions of constraints on the model (as well as derived) parameters. Given a likelihood function $L(\boldsymbol{\theta})$, with $\boldsymbol{\theta}=\{\theta_{\alpha}\}$ the set of model parameters, the Fisher matrix~\cite{1995PhDT........19B,Vogeley:1996xu,Tegmark:1996bz} is defined as
\begin{equation}
    F_{\alpha \beta} =\bigg\langle-\frac{\partial^{2} \ln L(\boldsymbol{\theta})}{\partial \theta_{\alpha} \partial \theta_{\beta}}\bigg |_\mathrm{fid}\bigg\rangle\,,
\end{equation}
where the angle brackets denote expectation values, and derivatives are computed at the reference fiducial parameters $\boldsymbol{\theta}_\mathrm{fid}$.

Assuming that $L(\boldsymbol{\theta})$ is a multivariate Gaussian distribution with respect to the model parameters, the Fisher matrix takes the form
\begin{equation}
    F_{\alpha \beta} =\frac{1}{2}\mathrm{tr}\left[\frac{\partial C_\mathrm{D}}{\partial\theta_\alpha}C_\mathrm{D}^{-1}\frac{\partial C_\mathrm{D}}{\partial\theta_\beta}C_\mathrm{D}^{-1}\right]+\sum_{ab}\frac{\partial \mu_a}{\partial\theta_\alpha}\left(C_\mathrm{D}^{-1}\right)_{ab}\frac{\partial \mu_b}{\partial\theta_\beta}\,,
\end{equation}
where $\boldsymbol{\mu}=\{\mu_{a}\}$ is the mean of the data vector $\boldsymbol{d}=\{d_{a}\}$, and $C_\mathrm{D}\equiv\langle(\boldsymbol{d}-\boldsymbol{\mu})(\boldsymbol{d}-\boldsymbol{\mu})^T\rangle$ is the expected data covariance matrix. Assuming that the Universe is described by the given model with certain values of parameters, the Fisher matrix, with $C_{\alpha\beta}=\left(F^{-1}\right)_{\alpha\beta}$ the full error covariance matrix of the parameters, provides expected errors around those {\it fiducial} values, and therefore, an estimate of the ability of an experiment or a combination of experiments to constrain the model.

Additionally, we can forecast constraints on any derived parameters $\boldsymbol{\tilde{\theta}}$ that are functions of $\boldsymbol{\theta}$ by calculating the Jacobian matrix $\tilde{J}_{\alpha\beta}\equiv \partial \tilde{\theta}_\alpha/\partial \theta_\beta$. The Fisher matrix associated to the new set of parameters is given by
\begin{equation}
    \tilde{F}=J^T F J,\label{eq:JacobianTransformation}
\end{equation}
where $J\equiv \tilde{J}^{-1}$. The constraints on the parameters $\boldsymbol{\tilde{\theta}}$ are then obtained from $\tilde{F}$.

This Fisher procedure requires the knowledge of the specifications of the experiments one considers and the observables one uses in comparing the models' predictions to the data. In the next three subsections, we briefly describe the cosmological observables and the surveys that we use in our Fisher forecast analysis of $\alpha$-attractor models, before presenting the results of our forecasts. We refer the reader to Ref.~\cite{Blanchard:2019oqi} for a detailed description of these observables and the Fisher formalism we use in the present paper. 

\subsection{Galaxy clustering}\label{sec:GC}

A {\it galaxy clustering} (GC) survey probes the correlation among positions of galaxies in the sky at different redshifts, which are assumed to be biased tracers of the distribution of dark matter in the Universe. The Fourier transform of the two-point correlation function of the dark matter density contrast $\delta_\mathrm{M}(x,z)$ at redshift $z$, known as the dark matter power spectrum $P_\mathrm{M}(k,z)$, characterizes the distribution of dark matter in the Universe. In the linear regime, $P_\mathrm{M}(k,z)$ is related to the primordial power spectrum of curvature perturbations generated by inflation through~\cite{Eisenstein:1997jh}
\begin{equation}\label{eq:DMPS}
    P_\mathrm{M}(k, z)=D^{2}(z) T^{2}(k) k^{n_{s}-1} \mathcal{A}_{s}\,,
    \end{equation}
where the transfer function has been split into the normalized, scale-independent growth factor $D(z)$ and an scale-dependent factor $T(k)$, normalized so that $T\rightarrow1$ for $k\rightarrow0$; see, e.g., Ref.~\cite{Eisenstein:1997jh} for a fitting function that provides $T(k)$.

What we observe through galaxy surveys is the power spectrum of galaxies. The galaxy density contrast $\delta_\mathrm{g}$ is tied to that of dark matter through $\delta_\mathrm{g}=b(z) \delta_\mathrm{M}$, where $b(z)$ is the so-called linear galaxy bias. The observed power spectrum of galaxies is then given in terms of the dark matter power spectrum as \cite{Seo:2003pu,Seo:2007ns,Wang:2012bx}
\begin{equation}
        P_{\mathrm{obs}}(z,k,\mu;\boldsymbol{\theta})=P_{\mathrm{s}}(z)+\frac{d_{\mathrm{A}}^{2}(z)_{\mathrm{fid}} H(z)}{d_{\mathrm{A}}^{2}(z) H(z)_{\mathrm{fid}}} b^{2}(z)\left(1+\beta(z) \mu^{2}\right)^{2} P_{\mathrm{M}}(k, z)e^{-k^{2} \mu^{2}\left(\sigma_{z}^{2}(z)c^2/H^2(z)+\sigma_{v}^{2}(z)\right)}\,,\label{eq:P_obs}
\end{equation}
where $P_{\mathrm{s}}(z)$ is the shot noise, $\mu\equiv{\mathbf{\hat z}\cdot\mathbf{ \hat k}}$ is the cosine of the angle between the normalized 3-wavevector ${\mathbf k}$ and the line of sight, $d_\mathrm{A}(z)$ is the angular diameter distance, $c$ is the speed of light, and the subscript `$\mathrm{fid}$' means that the corresponding quantity is calculated at the reference fiducial cosmology. Moreover, the final exponential term corresponds to a damping in the line-of-sight determination of $z$ due to peculiar velocity dispersions $\sigma_v(z)$ and observational spectroscopic redshift errors $\sigma_z(z)$. Expression~(\ref{eq:P_obs}) shows that the observed power spectrum depends on quantities that are themselves functions of the cosmological parameters. In addition to the direct appearance of background quantities $H$ and $d_{\mathrm{A}}$ determined by the cosmological model and its parameters, these quantities affect the observed power spectrum also through the growth rate $f$ introduced in Section~\ref{sec:pert}. $f$ enters the expression~(\ref{eq:P_obs}) through both the dark matter power spectrum $P_{\mathrm{M}}(k, z)$ and the Kaiser term~\cite{Kaiser:1987qv} $(1+\beta(z)\mu^2)^2$, with $\beta(z)\equiv f(z)/b(z)$. 

A GC survey in a redshift slice (or redshift bin) $\Delta z$ at redshift $z$ covers a volume $V_\mathrm{survey}$, where the galaxy number density $n(z)$ is given by the survey's specifications. The survey provides information for Fourier modes only in a range $\left[k_{\min}, k_{\max}\right]$, which also depends on the survey's specifications, as only scales larger than the survey's resolution are accessible.

Considering a GC survey carried out for a redshift range discretized into $N_b$ redshift bins, we evaluate $P_{\mathrm{obs}}(z, k, \mu; \boldsymbol{\theta})$ and its derivatives at the centre $\bar z_{m}$ of each redshift bin $m$, and at the fiducial value for each of the $N_p$ cosmological parameters. While we may expect the Fisher matrix to be an $ N_p \times N_p$ matrix, it is in practice more complicated due to the presence of $b(z)$ and $P_s(z)$, which are in general unknown. In order to address this problem, we discretize $b(z)$ and $P_s(z)$ into $N_b$ redshift bins, assuming them to be mutually independent and considering them at each redshift bin as additional independent model parameters with some fiducial values.\footnote{The fiducial values of the bias are given by the survey's specifications, and the fiducial `extra' shot noise is set to $0$.} Thus our full Fisher matrix is of dimension $\left( N_p+2N_b \right) \times \left( N_p+2N_b \right)$, and we can in the end marginalize it over the $2N_b$ parameters that are irrelevant to our cosmological model.

Given the extended parameter set $\boldsymbol{\Theta}=\left\{\theta_{\alpha}, b_{m}, P_{s, m}\right\}$, where $\theta_\alpha$ are the cosmological parameters, $b_m\equiv b(\bar z_{m})$ and $P_{s,m}\equiv P_s(\bar z_{m})$, the Fisher matrix for a GC survey at redshift $z$ can be written as~\cite{Blanchard:2019oqi}
\begin{equation}
    \begin{aligned}
    F_{\alpha\beta}=\frac{1}{8 \pi^{2}} \int_{-1}^{+1} \mathrm{d} \mu \int_{k_{\min }}^{k_{\max }} k^2 \mathrm{d} k \frac{\partial \ln P_{\mathrm{obs}}}{\partial \Theta_{\alpha}} \frac{\partial \ln P_{\mathrm{obs}}}{\partial \Theta_{\beta}} V_{\mathrm{eff}}\,,\label{eq:GCFisher}
    \end{aligned}
    \end{equation}
where
\begin{equation}
    V_{\mathrm{eff}}=V_{\text {survey}}\left[\frac{n(z) P_{\mathrm{obs}}}{n(z) P_{\mathrm{obs}}+1}\right]^{2}
    \end{equation}
is the effective volume of the survey.

The power spectrum and its derivatives appearing in Eq.~(\ref{eq:GCFisher}) are evaluated at their fiducial values and the final Fisher matrix is the combination of Fisher matrices at different redshift bins, i.e., the sum of the $N_b$ Fisher matrices. We also marginalize over the irrelevant parameters at this stage to obtain a matrix of dimension $ N_p \times N_p$ as the resulting $F^{\mathrm{GC}}_{\alpha\beta}$ for the cosmological parameters, which contains the constraint information about the parameter set $\boldsymbol{\theta}$.

\subsection{Weak lensing}\label{sec:WL}

A {\it weak lensing} (WL) survey probes gravitational lensing of the light emitted from distant galaxies due to the distribution of matter along the line of sight by measuring distortions in the shapes of the galaxies. WL surveys measure, simultaneously, both the geometry of the Universe and the growth of structure through the matter power spectrum. By measuring the correlations in the image distortions of galaxies, one can reconstruct the matter density field. This is done (see, e.g., Ref.~\cite{Blanchard:2019oqi}) through the measurement of the shear angular power spectrum
\begin{equation}\label{eq:cijdef}
C_{ij}^{\gamma\gamma}(\ell) = \frac{c}{H_0} 
\int{\frac{{\hat{W}}_i^\gamma(z) {\hat{W}}_{j}^\gamma(z)}{E(z) r^2(z)}
P_{\Phi+\Psi}\left ( k_{\ell}, z \right ) dz}\,,
\end{equation}
where $E(z)\equiv H(z)/H_0$, $r(z)$ is the comoving distance to the source at redshift $z$, and $P_{\Phi+\Psi}$ is the power spectrum of the Weyl potential $\Phi+\Psi$, with $\Phi$ and $\Psi$ the two Bardeen potentials. Here, the Limber approximation~\cite{Kaiser:1991qi,LoVerde:2008re,Giannantonio:2011ya,Kitching:2016zkn,Kilbinger:2017lvu,Lemos:2017arq} has been used to relate a wavenumber $k$ and a multipole $\ell$ through $k_\ell=(\ell+1/2)/r(z)$ (see, e.g., Ref.~\cite{Taylor:2018qda} for the full and exact computation). The indices $i$ and $j$ denote the redshift bins of a tomographic WL survey, allowing us to use the information on the time evolution of $\Phi+\Psi$ provided by the survey. Finally, the quantity $\hat{W}_{i}^\gamma(z)$ is related to the so-called lensing kernel, a purely geometrical quantity given by
\begin{equation}\label{eq:lenskernel}
\widetilde{W}_{i}^\gamma(z) = 
\int_{z}^{z_\text{max}}{n_{i}(z^{\prime}) \frac{r(z^{\prime}-z)}{r(z^{\prime})} dz^{\prime}}\,,
\end{equation}
through
\begin{equation}\label{eq:hatw}
\hat{W}_{i}^\gamma(z) = \frac{r(z)}{2}\widetilde{W}_{i}^\gamma(z)\,,
\end{equation}
where $n_i(z)$ is the normalized observed galaxy number density in the $i$th redshift bin.

In general-relativity-based cosmological models, including our $\alpha$-attractor models of quintessential inflation, the WL power spectrum $P_{\Phi+\Psi}$ is related to the matter power spectrum $P_\mathrm{M}$ by
\begin{equation}\label{eq:weylLAM}
    P_{\Phi+\Psi} =  \left[3\left(\frac{H_0}{c}\right)^2\Omega_\mathrm{M}^{0} (1 + z)\right]^2 P_\mathrm{M}\,.
\end{equation}
Combining Eqs.~(\ref{eq:cijdef}) and (\ref{eq:weylLAM}) leads to
\begin{equation}\label{eq:cijmatter}
C_{ij}^{\gamma\gamma}(\ell) = \frac{c}{H_0} 
\int{\frac{W_i^\gamma(z) W_{j}^\gamma(z)}{E(z) r^2(z)}
P_\text{M}\left ( k_{\ell}, z \right ) dz}\,,    
\end{equation}
where the window function $W_i^{\gamma}(z)$ is given by
\begin{equation}
    W_i^{\gamma}(z)= \left[3\left(\frac{H_0}{c}\right)^2\Omega_\mathrm{M}^{0} (1 + z)\right]\frac{r(z)}{2} 
\widetilde{W}_{i}^{\rm \gamma}(z)\,.
\end{equation}

We can now write the WL Fisher matrix (assuming the covariance to be the signal) as a sum over all multipoles correlating the signal at all redshift bins~\cite{Tegmark:1997rp,Carron:2012pw},
\begin{equation}
F_{\alpha\beta}=\sum_{\ell=\ell_{\rm min}}^{\ell_{\rm max}}\sum_{i,j,k,m}
\frac{\partial \widetilde{C}^{\gamma\gamma}_{ij}}{\partial\theta_\alpha} \{[\Delta C^{\gamma\gamma}]^{-1}\}_{jk} \frac{\partial \widetilde{C}^{\gamma\gamma}_{km}}{\partial\theta_\beta} \{[\Delta C^{\gamma\gamma}]^{-1}\}_{mi} \,, \label{eq:FisherSum-WL-new}
\end{equation} 
with
\begin{equation}
\widetilde C^{\gamma\gamma}_{ij}(\ell)=C_{ij}^{\gamma\gamma}(\ell)+\delta_{ij}\gamma_{\rm int}^{2}\widetilde{N}(n_{\theta}, \mathcal{N}_\mathrm{bin})_{i}^{-1}(\ell)\,,
\end{equation}
an estimation of the measured shear angular power spectrum for the fiducial cosmology with the theoretical power spectrum $C_{ij}^{\gamma\gamma}(\ell)$ and an expected noise $N_{ij}^{\gamma\gamma}(\ell)$ (the second term on the right-hand side of the equation). Here, $\gamma_{\rm int}$ is the intrinsic galaxy ellipticity and $\widetilde{N}_{i}^{-1}$ is a shot noise term for the $i$th redshift bin. This shot noise term depends on the total number of galaxies per square arcminute, $n_{\theta}$, and the total number of redshift bins $\mathcal{N}_\mathrm{bin}$ (for details, see Refs.~\cite{Casas:2015qpa,Casas:2017wjh,Amendola:2016saw}). The quantity 
\begin{equation}
\Delta C^{\gamma\gamma}_{ij}(\ell)=\sqrt{\frac{2}{(2\ell+1)\Delta\ell f_\mathrm{sky}}}\widetilde{C}^{\gamma\gamma}_{ij}(\ell)
\end{equation}
in Eq.~(\ref{eq:FisherSum-WL-new}) denotes the error on the observed shear angular power spectrum $\widetilde C^{\gamma\gamma}_{ij}(\ell)$, where $f_{\rm sky}$ is the fraction of surveyed sky and $\Delta\ell$ is the width of the multipole bins. We use $100$ logarithmically-spaced bins from $\ell_\mathrm{min}=10$ to $\ell_\mathrm{max}=1500$. The high-multipole cutoff $\ell_{\rm max}$ encodes our ignorance of clustering, systematics and baryon physics on small scales. 

\subsection{Stage IV galaxy surveys and specifications}\label{sec:GS-spec}

We consider three representative Stage IV galaxy surveys: LSST\footnote{\url{https://www.lsst.org}}~\cite{Ivezic:2008fe,Abell:2009aa,Mandelbaum:2018ouv}, DESI\footnote{\url{https://www.desi.lbl.gov}}~\cite{Aghamousa:2016zmz,Aghamousa:2016sne}, and SKA2, the second phase of the SKA\footnote{\url{https://www.skatelescope.org}}~\cite{Yahya:2014yva,Bacon:2018dui}.

LSST, with its telescope located in Chile and scheduled to start operations in 2021, will collect photometric data of billions of galaxies during an observation period of 10 years. It assumes a redshift distribution $dn/dz \propto (z/{z_0}_{p})^2 e^{-\left( z / z_0 \right)^\beta}$, with ${z_0}_{p}=1.0$, $z_0=0.11$ and $\beta=0.68$, truncated at redshift $z=4$ and normalized to $27$ galaxies per square arcminute in a total area of 14300 square degrees. The redshift sample is split into 10 tomographic bins with equal numbers of galaxies per bin. The fiducial total variance of the shape noise is set to $\sigma_\epsilon = 0.26$ and the expected relative photometric error is $0.05$. These specifications can be found in \cite{Mandelbaum:2018ouv,Abolfathi:2020can,Zhan:2017uwu,Ishak:2019aay}.

DESI is a ground-based optical telescope, located in Arizona and scheduled to start operations in 2021. It will study the formation and evolution of the large-scale structure by measuring spectroscopic redshifts and positions of 20-30 million galaxies and quasars with a relative spectroscopic redshift error of $10^{-3}$. It will cover an area of approximately 14000 square degrees. For the forecasts presented in this paper, we use the specifications for emission line galaxies (ELGs), as described in Refs.~\cite{Aghamousa:2016zmz,Flaugher:2014lfa}. The geometry and redshift binning specifications, as well as the galaxy number density and bias, can be found in Refs.~\cite{Aghamousa:2016zmz,Flaugher:2014lfa, Casas:2017eob}.

The SKA is an array of radio telescopes to be placed around the globe and built in two phases of SKA1 and SKA2. Here we consider SKA2~\cite{Yahya:2014yva}, which is classified as a Stage IV survey, and is scheduled to start operations in 2030. For the photometric WL probe we assume~\cite{Casas:2017eob,Harrison:2016stv} redshift distribution $dn/dz \propto (z/{z_0}_{p})^2 e^{-\left( z / z_0 \right)^\beta}$, with ${z_0}_{p}=z_0=1.1314$ and $\beta=1.5$, truncated at redshift $z=4$ and normalized to $10$ galaxies per square arcminute in a total area of 30940 square degrees. The redshift sample is split into 10 tomographic bins with equal numbers of galaxies per bin. The fiducial total variance of the shape noise is set to $\sigma_\epsilon = 0.3$ and the expected relative photometric error is $0.05$. In addition, we set $k_{\min}$ and $k_{\max}$ to $\SI{0.003}{\h\per\mega\parsec}$ and $\SI{0.20}{\h\per\mega\parsec}$, respectively. In this work, we only use the HI galaxy redshift survey mode of the SKA, and leave a full consideration of SKA2 for future work, where intensity mapping and continuum galaxy surveys will also be included---the required theoretical modeling tools for these surveys are currently unavailable.

In this paper, we present forecasts for cases where either only spectroscopic GC or a combination of spectroscopic GC and photometric WL has been used. For simplicity, we leave out photometric GC and its cross correlation with the photometric cosmic shear (WL) probe.
We assume the spectroscopic measurements of DESI to be independent of the photometric measurements of LSST, and therefore, we combine their forecasts by simply adding their individual Fisher matrices. SKA2 provides both spectroscopic GC and photometric WL data, based on different sources. Therefore, we can assume independent measurements also in this case, and combine them in the same way as described above.

\subsection{Fiducial parameters and Fisher matrix analysis}\label{sec:fid-Fisher}

In this section we present the results of our Fisher forecast analysis. This analysis is performed using \texttt{Fishermathica}\footnote{\url{https://github.com/santiagocasas/fishermathica}}~\cite{casas_castro_non-linear_2017}, a multipurpose \texttt{Wolfram Mathematica} code designed for Fisher analysis. \texttt{Fishermathica} works in cooperation with \texttt{Cosmomathica}\footnote{\url{https://github.com/santiagocasas/cosmomathica}}~\cite{Amendola:2013qna}, a \texttt{Wolfram Mathematica} wrapper for Einstein-Boltzmann codes \texttt{CAMB}~\cite{Lewis:1999bs} and \texttt{CLASS}~\cite{Lesgourgues:2011rg,Lesgourgues:2011re}, fitting function codes~\cite{Eisenstein:1997jh}, and \texttt{Halofit}~\cite{Smith:2002dz, Takahashi2012}.
\texttt{Fishermathica} has been extensively tested and validated within the \textit{Euclid} IST:Forecasts code comparison effort. See Ref.~\cite{Blanchard:2019oqi} for details of the code validation and testing the accuracy of the derivatives, which is the most challenging task when performing Fisher matrix forecasts.

\begin{table}
\small
\rowcolors{1}{}{lightgray}
    \centering
    \begin{tabular}{ c|c } 
     \rowcolor{blizzardblue} 
     Parameter & Fiducial value \\ 
     \hline
     \hline
     $\Omega_\mathrm{M}^{0}$ & 0.32 \\ 
     $\Omega_\mathrm{b}^{0}$ & 0.05 \\
     $\gamma$ (Exp-model I;~~$\alpha\!=\!7/3$, $\varphi_\mathrm{F}\!=\!-10$) & 127.871 \\
     $\gamma$ (Exp-model I;~~$\alpha\!=\!1/3$, $\varphi_\mathrm{F}\!=\!-35$) & 126.254 \\
     $\gamma$ (Exp-model II;~$\alpha\!=\!7/3$, $\varphi_\mathrm{F}\!=\!-10$) & 127.645 \\
     $\gamma$ (Exp-model II;~$\alpha\!=\!2/3$, $\varphi_\mathrm{F}\!=\!-10$) & 124.258 \\
     $n_s$ & \num{0.966} \\
     $\mathcal{A}_s$ & \num{2.12605e-9} \\
     \hline
     \end{tabular}
    \caption{Fiducial values for free parameters of exponential-potential $\alpha$-attractor models of quintessential inflation directly varied in the Fisher analysis of the present work. The four different fiducial values of $\gamma$ correspond to the four cases specified in the parentheses, and are obtained iteratively in order to maintain $h\approx 0.72$.}\label{table:fidu}
 \end{table}

The fiducial values of cosmologically relevant parameters are summarized in Table~\ref{table:fidu}. Note that the baryon fraction $\Omega_\mathrm{b}^{0}$ is included as it enters the transfer function $T(k)$. $\Omega_\mathrm{R}^{0}$, on the other hand, is fixed in \texttt{Fishermathica} and is not varied in our analysis. For $\Omega_\mathrm{M}^{0}$, $\Omega_\mathrm{b}^{0}$ and $\mathcal{A}_s$, we choose fiducial values consistent with existing constraints on these parameters based on $\Lambda$CDM fits, while for $n_s$ we choose the fiducial value of $0.966$, corresponding to an $N_*\sim60$, which is typical for quintessential $\alpha$-attractors. The fiducial value of $\gamma$ is iteratively computed from the requirement of having $H_0=\SI{72}{\kilo\meter\per\second\per\mega\parsec}$ (or $h=0.72$).\footnote{We do not expect our $\alpha$-attractor models to address existing tensions in the measured values of $H_0$~\cite{Verde:2019ivm}. We therefore set its fiducial value to $\SI{72}{\kilo\meter\per\second\per\mega\parsec}$ as a representative value. Note that the exact value of $H_0$ is not important in our forecast analysis, as we do not perform any parameter estimation based on the observed data, and the forecasted constraints we obtain on $H_0$ and other parameters do not depend on the chosen fiducial value for $H_0$. See, however, Ref.~\cite{Braglia:2020bym} for an application of $\alpha$-attractors in context of the Hubble tension.} This means that depending on the model and values of other parameters, the fiducial values of $\gamma$ are generically different in different cases. 

The parameter $\alpha$ is fixed to specific values, rather than  being varied as a free parameter. The reason for this is that, as we discussed in Section~\ref{sec:alphaattractors}, there are values of $\alpha$ that are of particular interest from the theoretical point of view. Therefore, we fix $\alpha$ to a value in the set of discrete values $\{1/3,2/3,1,4/3,5/3,2,7/3\}$. $\alpha=7/3$ is particularly interesting as it is one of the benchmark targets for future CMB $B$-mode polarization experiments with a detectably large signal~\cite{Kallosh:2019eeu,Kallosh:2019hzo}. Similarly, we also fix $\varphi_\mathrm{F}$ in our analysis. As shown in Ref.~\cite{Akrami:2017cir}, it can take values in the range $[-35,-10]$ for exponential potentials, depending on the reheating mechanism. The lower limit is set by gravitational reheating as the only reheating mechanism, and the upper limit comes from the viability condition for the cosmic history at late times when more efficient reheating mechanisms are at work---the lower the value of $|\varphi_\mathrm{F}|$, the more deviations from a $\Lambda$CDM universe. We therefore consider $\varphi_\mathrm{F}$ as a model classifier and fix it to either of the two values $-35$ and $-10$, for the smallest and largest deviations from $\Lambda$CDM, respectively. 

 \begin{table}
\small
\rowcolors{1}{}{lightgray}
    \centering
    \begin{tabular}{ c|c|c|c} 
     \rowcolor{blizzardblue} 
     Model & $w_0$ & $w_a$ & $M^2[M_\mathrm{Pl}^2]$\\ 
     \hline
     \hline
     Exp-model I;~~$\alpha\!=\!7/3$, $\varphi_\mathrm{F}\!=\!-10$ & $-0.9457$ & $-0.0631$ & \num{3.06157e-10}\\
     Exp-model I;~~$\alpha\!=\!1/3$, $\varphi_\mathrm{F}\!=\!-35$ & $\sim -1$ & $\sim 0$ & \num{3.73288e-11}   \\
     Exp-model II;~$\alpha\!=\!7/3$, $\varphi_\mathrm{F}\!=\!-10$ & $-0.8849$ & $-0.1344$ & \num{3.06157e-10}  \\
     Exp-model II;~$\alpha\!=\!2/3$, $\varphi_\mathrm{F}\!=\!-10$ & $-0.8490$ & $-0.1927$ & \num{7.66111e-11} \\
     \hline
     \end{tabular}
    \caption{Present-day values and time derivatives, $w_0$ and $w_a$, of dark energy equation of state, as well as the values of the mass scale $M^2$, corresponding to the fiducial models presented in Table~\ref{table:fidu}.}\label{table:fidu-derived}
 \end{table}
 
As summarized in Table~\ref{table:fidu}, we consider four specific fiducial models, with different fixed values of $\alpha$ and $\varphi_\mathrm{F}$---we discuss this choice of fiducial models shortly. We also provide in Table~\ref{table:fidu-derived} the corresponding values of the present-day dark energy equation of state and its time derivative, $w_0$ and $w_a$, as well as the mass scale of the potential, $M^2$, in Planck units and computed from Eq.~(\ref{eq:cobe}). 

\begin{figure*}
    \centering
    \includegraphics[scale=0.55]{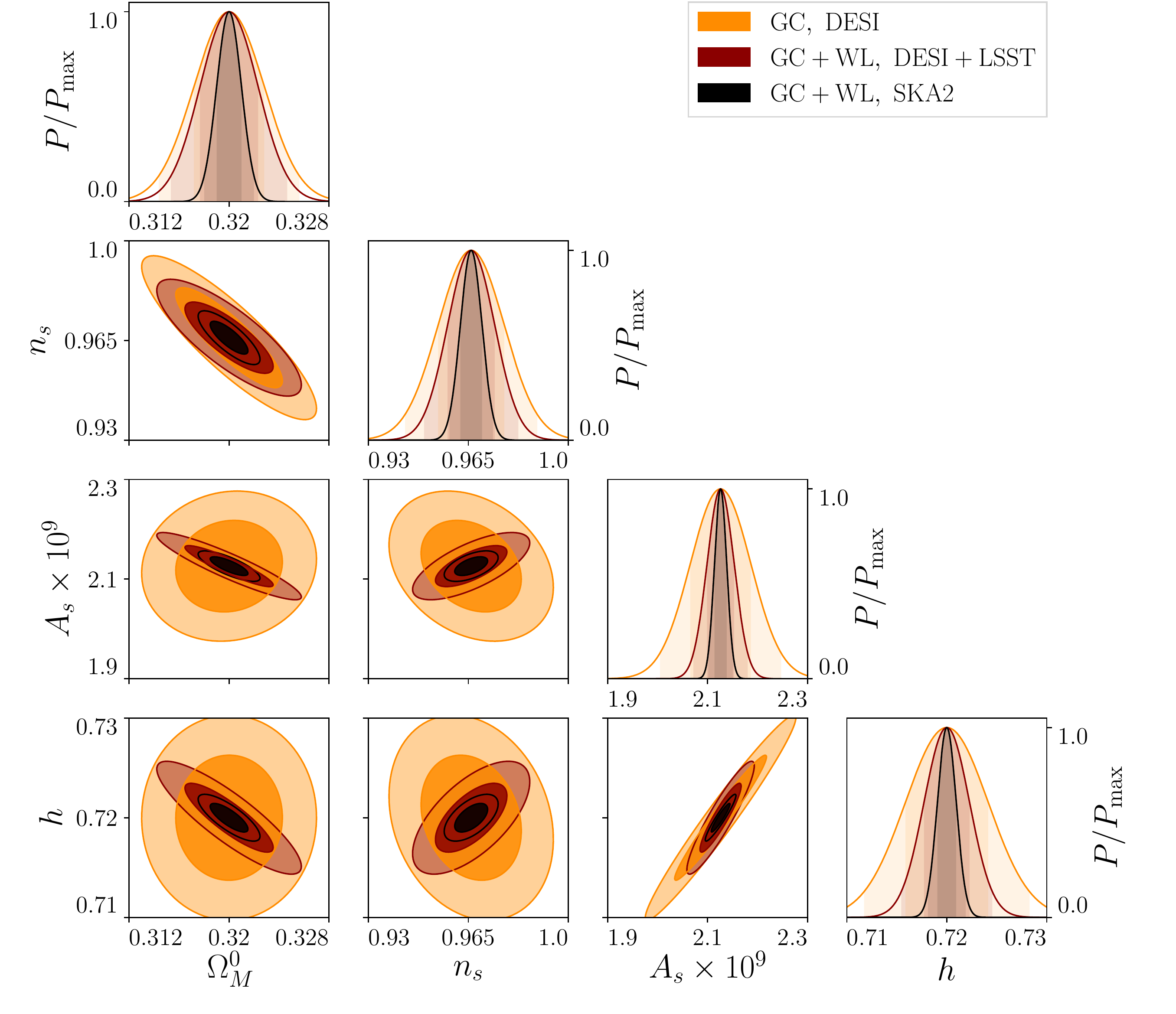}
    \caption{Fisher-matrix $68.3\%$ and $95\%$ confidence regions and 1-dimensional marginalized probability density functions for the standard $\Lambda$CDM model. For each panel, we show the results for both cases of galaxy clustering (GC) alone, provided by DESI, and the combination of galaxy clustering and weak lensing (GC+WL), provided by either DESI in combination with LSST or SKA2.}\label{fig:DESI-LSST-SKA2-LCDM_triplot_ellipses}
\end{figure*}

In order to validate our analysis, we first consider the $\Lambda$CDM model with the same fiducial values for its parameters $\{\Omega_\mathrm{M}^{0}, \Omega_\mathrm{b}^{0}, n_s, \mathcal{A}_s, h\}$ as those we have set for the parameters of our $\alpha$-attractor models. In Fig.~\ref{fig:DESI-LSST-SKA2-LCDM_triplot_ellipses}, we show the obtained $68.3\%$ and $95\%$ confidence constraints on four of these parameters, i.e., $\Omega_\mathrm{M}^{0}$, $n_s$, $\mathcal{A}_s$ and $h$, when either only GC or a combination of GC and WL is considered. We present our results for DESI, DESI+LSST and SKA2; see also Table~\ref{table:constraints} for $68.3\%$ (or $1\sigma$) uncertainties on these parameters for DESI+LSST and SKA2, where combinations of GC and WL surveys are considered. Fig.~\ref{fig:DESI-LSST-SKA2-LCDM_triplot_ellipses} illustrates that in all cases, combining CG and WL tightens the constraints on all parameters, as expected. Additionally, we observe that the SKA2 constraints are significantly stronger than those from the combination of DESI and LSST. This is expected, as there are several differences between SKA2 and the other surveys. For example, the area of the sky covered by SKA2 is larger than those covered by DESI and LSST by a factor of $> 2$. Additionally, SKA2 covers a significantly larger range of redshifts. This can be seen from the redshift distributions provided in Section~\ref{sec:GS-spec}.

\begin{figure*}
    \centering
    \includegraphics[scale=0.55]{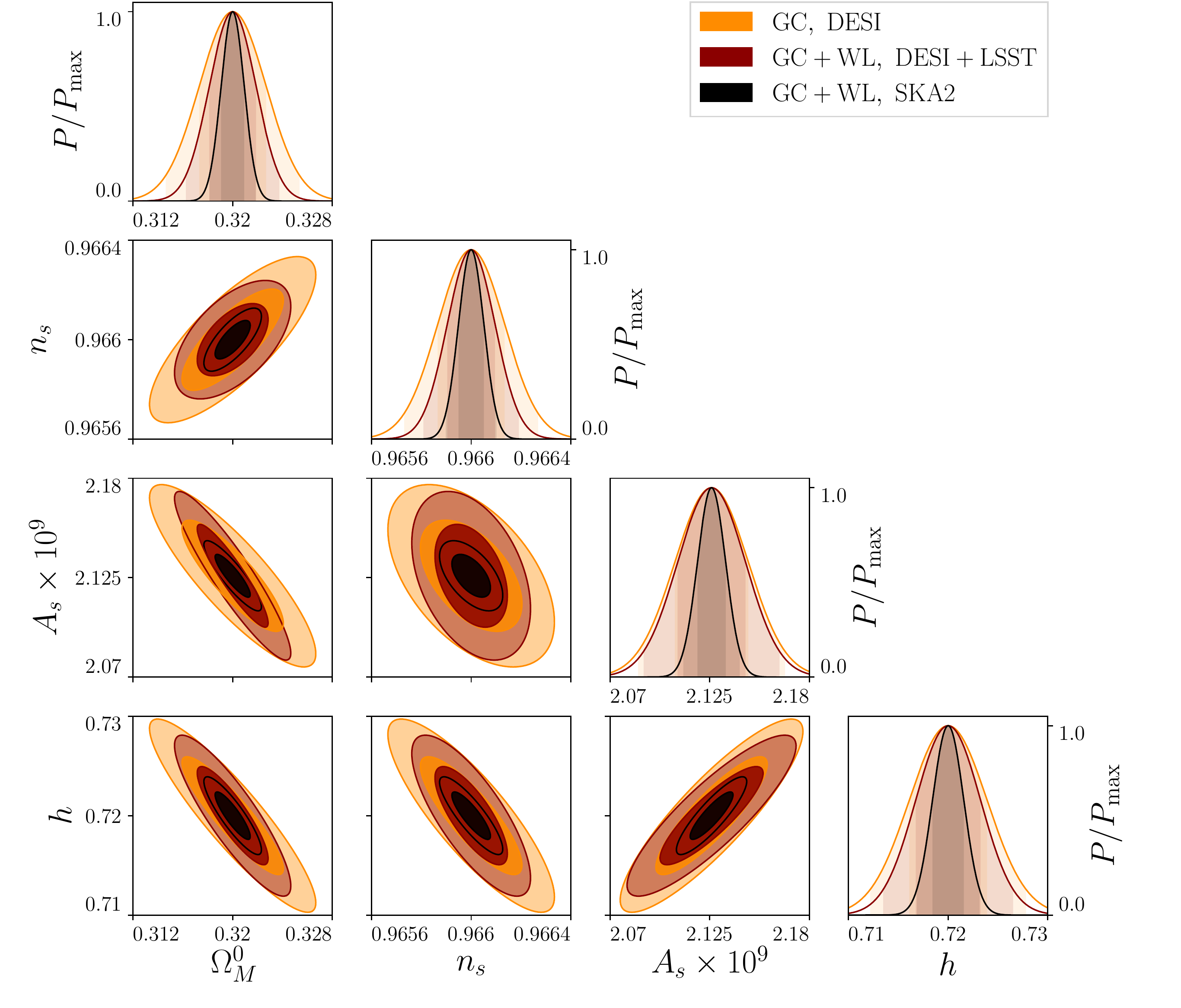}
    \caption{As in Fig.~\ref{fig:DESI-LSST-SKA2-LCDM_triplot_ellipses}, but for the $\alpha$-attractor quintessential inflation model Exp-model I with $\alpha=7/3$ and $\varphi_\mathrm{F}=-10$. Here, $\Omega_\mathrm{M}^{0}$, $n_s$ and $\mathcal{A}_s$ are primary parameters varied through the Fisher forecast analysis, while $h$ is a derived parameter.}\label{fig:DESI-LSST-SKA2-ExpI_triplot_ellipses}
\end{figure*}

Turning now to our $\alpha$-attractor models, we first focus on Exp-model I, and in Fig.~\ref{fig:DESI-LSST-SKA2-ExpI_triplot_ellipses}, we present the results of our Fisher forecast analysis for the case with $\alpha=7/3$ and $\varphi_\mathrm{F}=-10$. As we discussed in Section~\ref{sec:mod-params} based on Ref.~\cite{Akrami:2017cir}, as well as Figs.~\ref{fig:wDE},~\ref{fig:H} and~\ref{fig:f}, this case is expected to provide the largest deviation of Exp-model I from $\Lambda$CDM in the entire ranges of $\alpha$ and $\varphi_\mathrm{F}$ values considered in our studies. Note that we have shown the results in Fig.~\ref{fig:DESI-LSST-SKA2-ExpI_triplot_ellipses} for the exact same set of parameters as in $\Lambda$CDM, i.e., $\{\Omega_\mathrm{M}^{0}, n_s, \mathcal{A}_s, h\}$. The parameters $\Omega_\mathrm{M}^{0}$, $n_s$ and $\mathcal{A}_s$ are three of the primary parameters that we varied through the Fisher analysis, while $h$ is a derived parameter. The reason why we show our results in terms of this specific set of parameters is that it makes it easier to compare the constraints with those on the $\Lambda$CDM model. Table~\ref{table:constraints} reports $1\sigma$ uncertainties on the parameters $\Omega_\mathrm{M}^{0}$, $n_s$, $\mathcal{A}_s$ and $h$ when combinations of GC and WL surveys are used, i.e., for DESI+LSST and SKA2.

 \begin{table}
 \footnotesize
    \centering
    \setlength\tabcolsep{4pt}
    \begin{tabular}{ l|c|c|c|c|c|c|c|c }
    \rowcolor{blizzardblue} 
     Data set/Model & $\Omega_\mathrm{M}^{0}$ & $n_s$ & $\mathcal{A}_s\!\times\!10^9$ & $h$ & $w_0$ & $w_a$ & $\gamma$ & $M^2\!\times\!10^{10}$ \\
     \hline
     \hline
     \rowcolor{blizzardblue} 
     \multicolumn{9}{c}{$\Lambda$CDM}\\
     \hline
     \hline
       \shaderow DESI+LSST & 0.002 & \num{0.008} & 0.027 & 0.002 & - & - & - & - \\ 
       SKA2 & 0.001 & \num{0.004} & 0.012 & 0.001 & - & - & - & - \\
     \hline
     \hline
     \rowcolor{blizzardblue} 
     \multicolumn{9}{c}{Exp-model I}\\
     \hline
     \hline
     \shaderow $\alpha\!=\!7/3$, $\varphi_\mathrm{F}\!=\!-10$, DESI+LSST & 0.002 & \num{0.0001} & 0.019 & 0.003 & 0.0002 & 0.00013 & 0.0006 & 0.040 \\
     $\alpha\!=\!7/3$, $\varphi_\mathrm{F}\!=\!-10$, SKA2 & 0.001 & \num{0.0001} & 0.008 & 0.002 & 0.0001 & 0.00006 & 0.0003 & 0.020 \\
     \shaderow $\alpha\!=\!1/3$, $\varphi_\mathrm{F}\!=\!-35$, DESI+LSST & 0.002 & \num{0.0001} & 0.018 & 0.003 & - & - & 0.0006 & 0.005\\
     $\alpha\!=\!1/3$, $\varphi_\mathrm{F}\!=\!-35$, SKA2 & 0.001 & \num{0.0001} & 0.008 & 0.002 & - & - & 0.0003 & 0.002\\
     \hline
     \hline
     \rowcolor{blizzardblue} 
     \multicolumn{9}{c}{Exp-model II}\\
     \hline
     \hline
     \shaderow $\alpha\!=\!7/3$, $\varphi_\mathrm{F}\!=-\!10$, DESI+LSST & 0.002 & \num{0.0001} & 0.018 & 0.003 & 0.00044 & 0.00020 & 0.0006 & 0.040 \\
     $\alpha\!=\!7/3$, $\varphi_\mathrm{F}\!=\!-10$, SKA2 & 0.001 & \num{0.0001} & 0.008 & 0.002 & 0.00018 & 0.00010 & 0.0003 & 0.020 \\
     \shaderow $\alpha\!=\!2/3$, $\varphi_\mathrm{F}\!=\!-10$, DESI+LSST & 0.002 & \num{0.0001} & 0.019 & 0.003 & 0.00060 & 0.00004 & 0.0006 & 0.010\\
     $\alpha\!=\!2/3$, $\varphi_\mathrm{F}\!=\!-10$, SKA2 & 0.001 & \num{0.0001} & 0.008 & 0.002 & 0.00034 & 0.00002 & 0.0003 & 0.005\\
     \hline
    \end{tabular}
    \caption{Fisher-matrix $1\sigma$ uncertainties on parameters $\Omega_\mathrm{M}^{0}$, $n_s$, $\mathcal{A}_s$ and $h$ for $\Lambda$CDM and $\alpha$-attractor quintessential inflation models Exp-model I and Exp-model II, as well as on $\alpha$-attractor-specific parameters $w_0$, $w_a$, $\gamma$ and $M^2$ (in units of Planck mass squared), when combinations of both galaxy clustering (GC) and weak lensing (WL) surveys are used. The results are provided for the combination of DESI and LSST, as well as SKA2.} \label{table:constraints}
 \end{table}
 
\begin{figure*}
    \centering
    \includegraphics[scale=0.55]{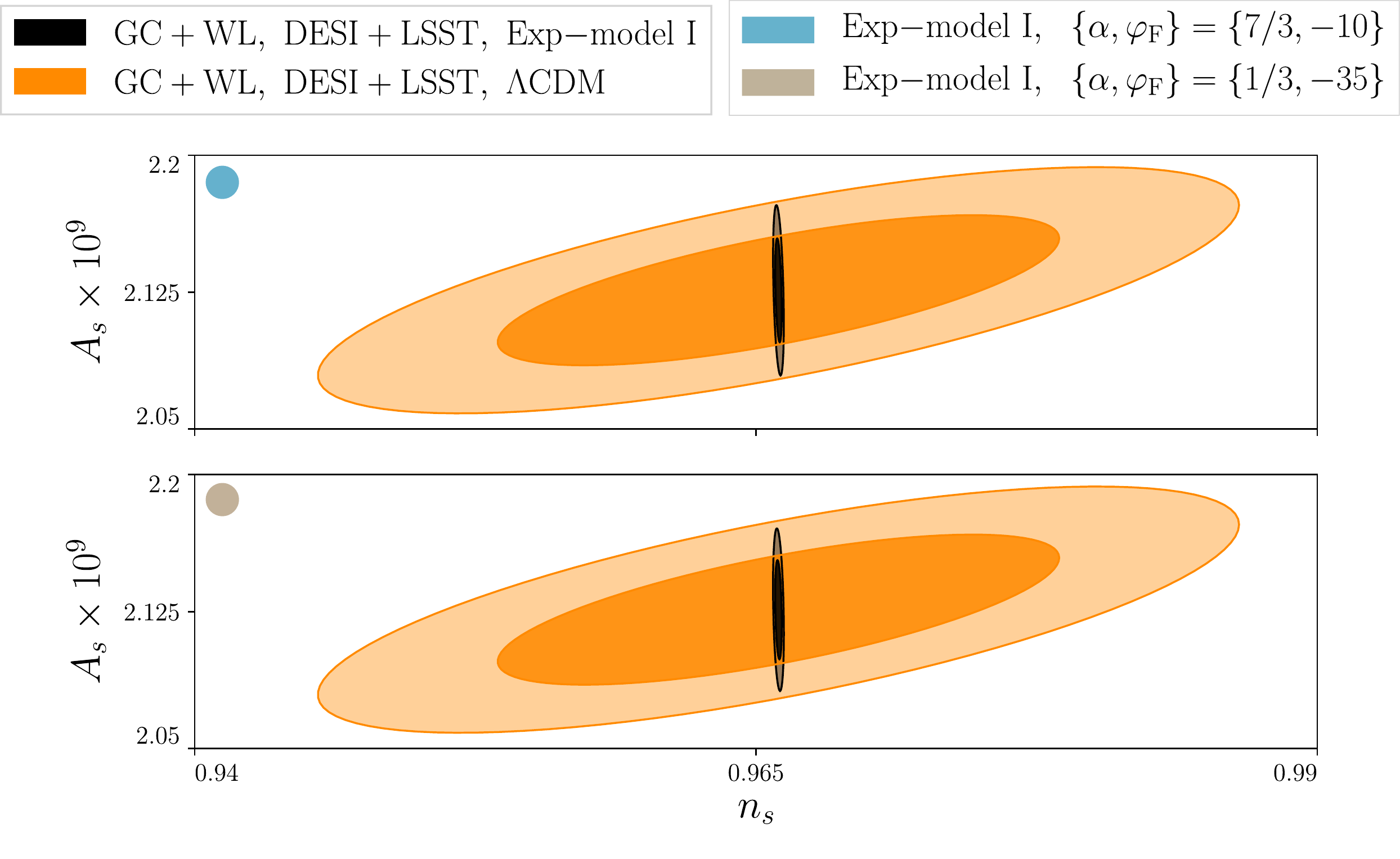}
    \caption{Fisher-matrix $68.3\%$ and $95\%$ confidence constraints on the spectral index and amplitude of primordial scalar perturbations, $n_s$ and $\mathcal{A}_s$, for the standard $\Lambda$CDM model versus the $\alpha$-attractor quintessential inflation model Exp-model I with $\{\alpha, \varphi_\mathrm{F}\}=\{7/3,-10\}$ (upper panel) and $\{\alpha,\varphi_\mathrm{F}\}=\{1/3,-35\}$ (lower panel). For each panel, we show the results for the combination of galaxy clustering (GC) and weak lensing (WL) to be measured by DESI and LSST.}\label{fig:DESI-LSST-LCDM-ExpI_As-ns_ellipses}
\end{figure*}

Both Fig.~\ref{fig:DESI-LSST-SKA2-ExpI_triplot_ellipses} and Table~\ref{table:constraints} show that the conclusions we made earlier for $\Lambda$CDM in terms of the constraints from DESI+LSST versus SKA2 and from GC versus GC+WL hold also for Exp-model I. However, the main difference between the two cases of $\alpha$-attractor and $\Lambda$CDM models is the much tighter constraints on the scalar spectral index $n_s$. To see this more clearly, we show in Fig.~\ref{fig:DESI-LSST-LCDM-ExpI_As-ns_ellipses} the 2-dimensional Fisher constraints from DESI+LSST on $\mathcal{A}_s$ and $n_s$ for Exp-model I and $\Lambda$CDM, where combinations of GC and WL surveys are used.

It is also interesting to know how these constraints change if we set $\alpha$ and $\varphi_\mathrm{F}$ to other values. A natural choice is $\{\alpha,\varphi_\mathrm{F}\}=\{1/3,-35\}$, which results in a cosmology closest to $\Lambda$CDM within the parameter space of our Exp-model I. Our analysis shows that the constraints on all parameters $\{\Omega_\mathrm{M}^{0}, n_s, \mathcal{A}_s, h\}$ remain very similar to the ones shown in Fig.~\ref{fig:DESI-LSST-SKA2-ExpI_triplot_ellipses} for the previous case of $\{\alpha, \varphi_\mathrm{F}\}=\{7/3,-10\}$. Therefore, for brevity, we do not present all the constraint plots for $\{\alpha,\varphi_\mathrm{F}\}=\{1/3,-35\}$ and only report in Table~\ref{table:constraints} the $1\sigma$ uncertainties on the parameters for the combination of GC and WL measurements to be made by DESI+LSST and SKA2. Additionally, we show in Fig.~\ref{fig:DESI-LSST-LCDM-ExpI_As-ns_ellipses} the 2-dimensional Fisher constraints on $\mathcal{A}_s$ and $n_s$ compared to $\Lambda$CDM. Since the two cases shown in the figure correspond to the highest and lowest deviations from $\Lambda$CDM for Exp-model I, we can conclude that the spectral index $n_s$ for any other choices of $\alpha\in[1/3,7/3]$ and $\varphi_\mathrm{F}\in[-35,-10]$ will also be highly constrained.

In order to understand the tight constraints on $n_s$ in $\alpha$-attractor quintessential inflation compared to the standard model, let us recall that $n_s$ in $\Lambda$CDM only enters the expression for the primordial power spectrum of curvature perturbations. In other words, it specifies the initial seed of the large-scale structure, but has no impacts on how the modes evolve subsequently. This means that when, for example, galaxy clustering is used to measure the expansion history and growth of structure, $n_s$ affects the observed power spectrum of galaxies, Eqs.~(\ref{eq:P_obs}), only through the factor $k^{n_s-1}$ in Eq.~(\ref{eq:DMPS}) for the dark matter power spectrum; this gives the $\Lambda$CDM constraints on $n_s$. The story is different for $\alpha$-attractor models of quintessential inflation. Eq.~(\ref{eq:cobe}) tells us that the value of $M$ is now affected directly by the value of $n_s$ (through the relation between $n_s$ and $N_*$). $M$ is a quantity that appears in the dynamical equations for the late-time evolution of the Universe, and therefore, its value affects the observed power spectrum through quantities like the Hubble expansion rate $H(z)$. As a result, $n_s$ is highly constrained in our $\alpha$-attractor models compared to $\Lambda$CDM or any other models where inflation is non-quintessential, determining only the initial conditions of the Universe.

\begin{figure*}
    \centering
    \includegraphics[scale=0.6]{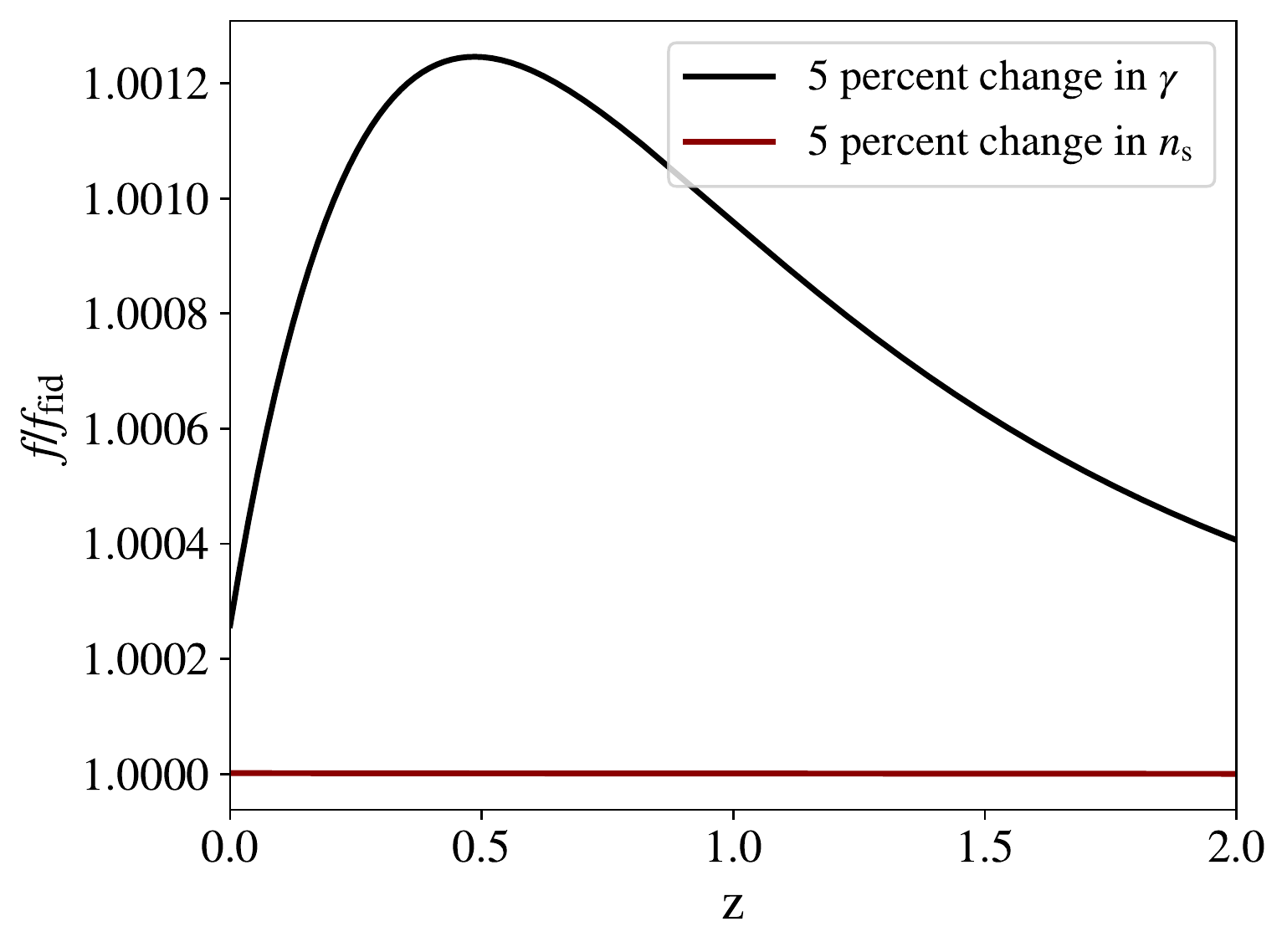}
    \caption{Changes in the growth rate $f(z)$ when parameters $n_s$ and $\gamma$ are changed by $5\%$ compared to their fiducial values.}\label{fig:degeneracy}
\end{figure*}

One may then ask what is the role of the parameter $\gamma$ here, and why can we not compensate for the tight constraints on $n_s$ by changing the value of $\gamma$, which is a parameter that together with $n_s$ (through $M$) determines the late-time dynamics? At the level of the background dynamics, $n_s$ and $\gamma$ are indeed degenerate. For example, reducing $n_s$ by $5\%$ leads to a change in the value of $H_0$ given by $H^\mathrm{changed}_0/H^\mathrm{fiducial}_0 \approx 2.79$, while reducing $\gamma$ by the same percentage leads to $H^\mathrm{changed}_0/H^\mathrm{fiducial}_0 \approx 500$. This means that by changing these parameters in opposite directions, we can keep the background unmodified. The degeneracy is, however, broken through changes in the growth rate $f(z)$. This is illustrated in Fig.~\ref{fig:degeneracy}, where we show changes in $f(z)$ when $n_s$ and $\gamma$ are changed by $5\%$ compared to their fiducial values. As we see, $f(z)$ is nearly insensitive to changes in $n_s$ while being sensitive to changes in $\gamma$.  Since $\gamma$ has the additional impact on observables such as the observed power spectrum of galaxies due to its effect on $f(z)$, it is tightly constrained, the degeneracy between $n_s$ and $\gamma$ is broken, and $n_s$ is highly constrained in $\alpha$-attractor quintessential inflation compared to $\Lambda$CDM, where $n_s$ does not affect $H_0$.

\begin{figure*}
    \centering
    \includegraphics[scale=0.7]{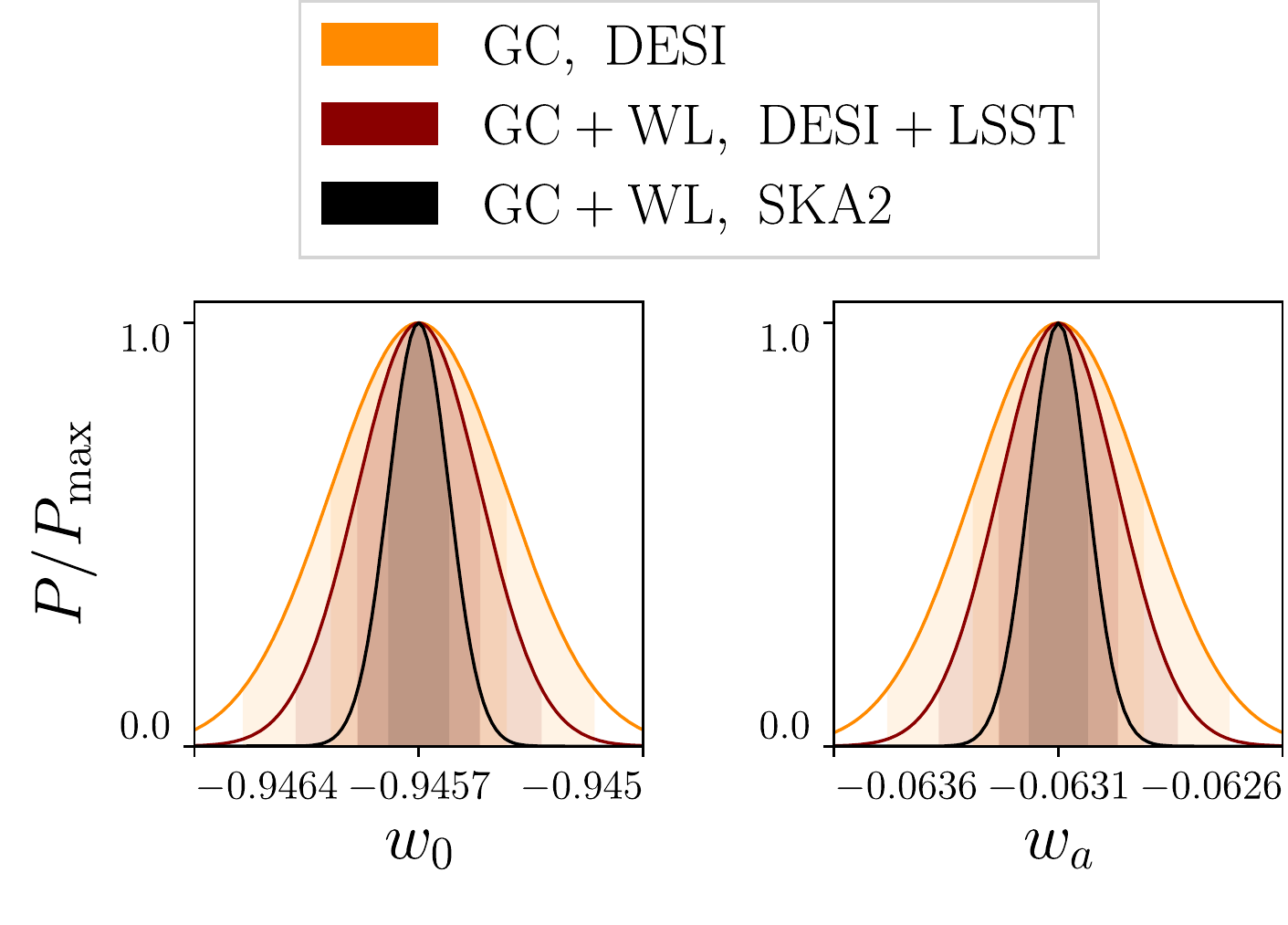}
    \caption{Fisher-matrix 1-dimensional marginalized probability density functions for the CPL parameters $w_0$ and $w_a$ and for the $\alpha$-attractor quintessential inflation model Exp-model I with $\{\alpha, \varphi_\mathrm{F}\}=\{7/3,-10\}$. For each panel, we show the results for galaxy clustering (GC) alone, provided by DESI, and the combination of galaxy clustering and weak lensing (GC+WL), provided by either DESI in combination with the LSST or SKA2.}\label{fig:DESI-LSST-SKA2-ExpI_1Dplot_w0wa}
\end{figure*}

Now that all the parameters we have discussed so far are similarly constrained by observational data for both cases of $\{\alpha,\varphi_\mathrm{F}\}=\{1/3,-35\}$ and $\{\alpha, \varphi_\mathrm{F}\}=\{7/3,-10\}$, one may wonder how to see the differences between these cases at the level of the cosmological parameters. One particularly important place where this difference can clearly be seen is the constraints on the derived CPL parameters $w_0$ and $w_a$ for the dark energy equation of state. Since Exp-model I with $\{\alpha,\varphi_\mathrm{F}\}=\{1/3,-35\}$ provides late-time dynamics that are extremely close to that of $\Lambda$CDM, the fiducial values of $w_0$ and $w_a$ are extremely close to $-1$ and $0$, respectively. The forecasted constraints on $w_0$ and $w_a$ for the $\{\alpha, \varphi_\mathrm{F}\}=\{7/3,-10\}$ case are shown in Fig.~\ref{fig:DESI-LSST-SKA2-ExpI_1Dplot_w0wa}, where we see tight constraints on both parameters, distinguishing them from the $\Lambda$CDM values. Table~\ref{table:constraints} provides $1\sigma$ uncertainties on the these parameters for the combination GC+WL to be measured by DESI+LSST and SKA2.

We now present our results for Exp-model II, and first analyze the model with $\alpha=7/3$, i.e., the largest $\alpha$ in our set of theoretically motivated values $\{1/3,2/3,1,4/3,5/3,2,7/3\}$. As we stated earlier, $\alpha=7/3$ corresponds to one of the benchmark targets for future CMB $B$-mode polarization experiments as it provides the largest and most detectable value of the tensor-to-scalar ratio $r$ in the range of $\alpha$ values considered in this paper~\cite{Kallosh:2019eeu,Kallosh:2019hzo}. Additionally, in order to maximize the deviation from $\Lambda$CDM, we fix $\varphi_\mathrm{F}$ to the largest value of $-10$; recall that $\varphi_\mathrm{F}=-10$ corresponds to a scenario with the most efficient reheating mechanism while currently being observationally viable~\cite{Akrami:2017cir}.

\begin{figure*}
    \centering
    \includegraphics[scale=0.55]{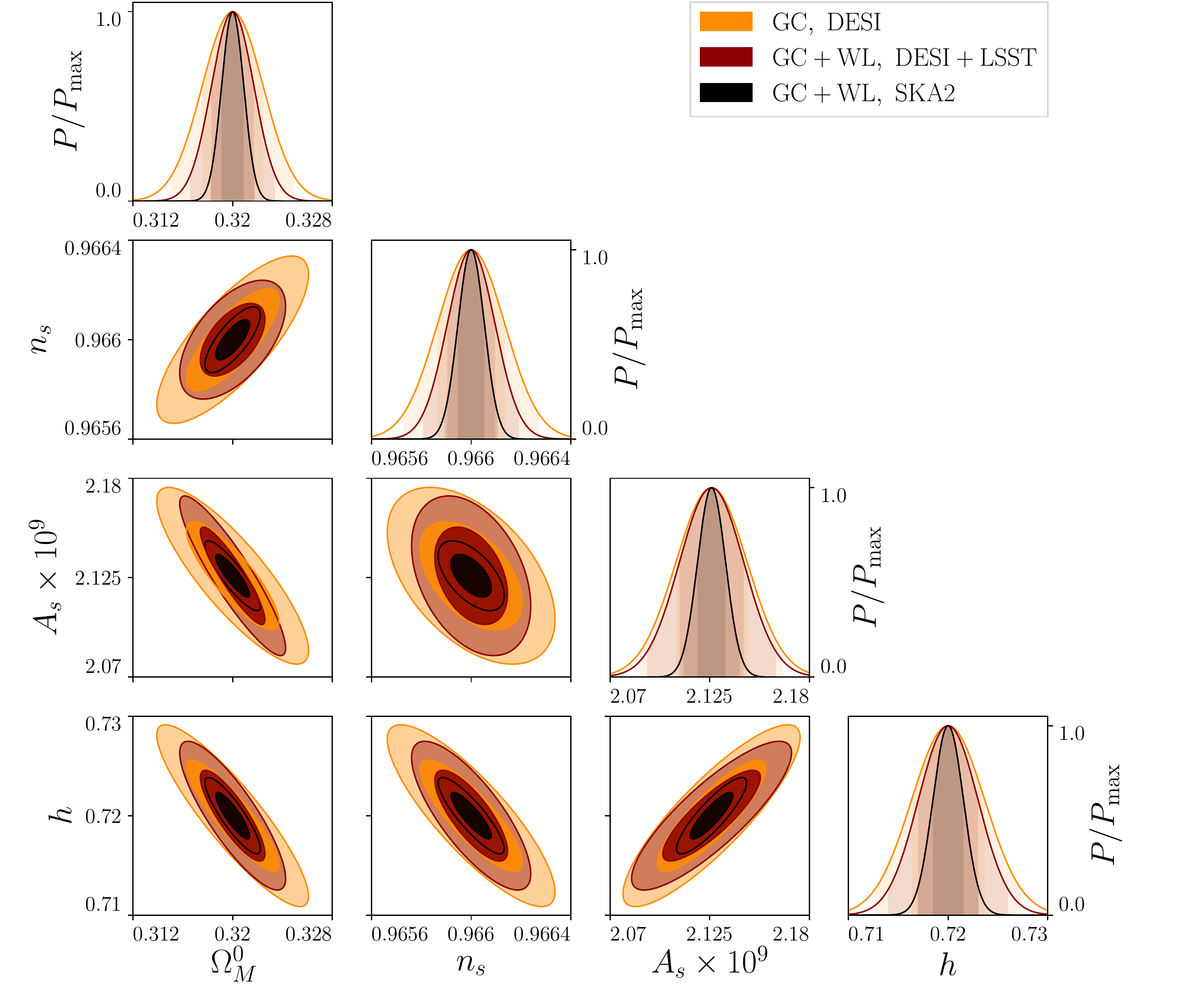}
    \caption{As in Fig.~\ref{fig:DESI-LSST-SKA2-ExpI_triplot_ellipses}, but for the $\alpha$-attractor quintessential inflation model Exp-model II with $\alpha=7/3$ and $\varphi_\mathrm{F}=-10$.}\label{fig:DESI-LSST-SKA2-ExpII_triplot_ellipses}
\end{figure*}

\begin{figure*}
    \centering
    \includegraphics[scale=0.55]{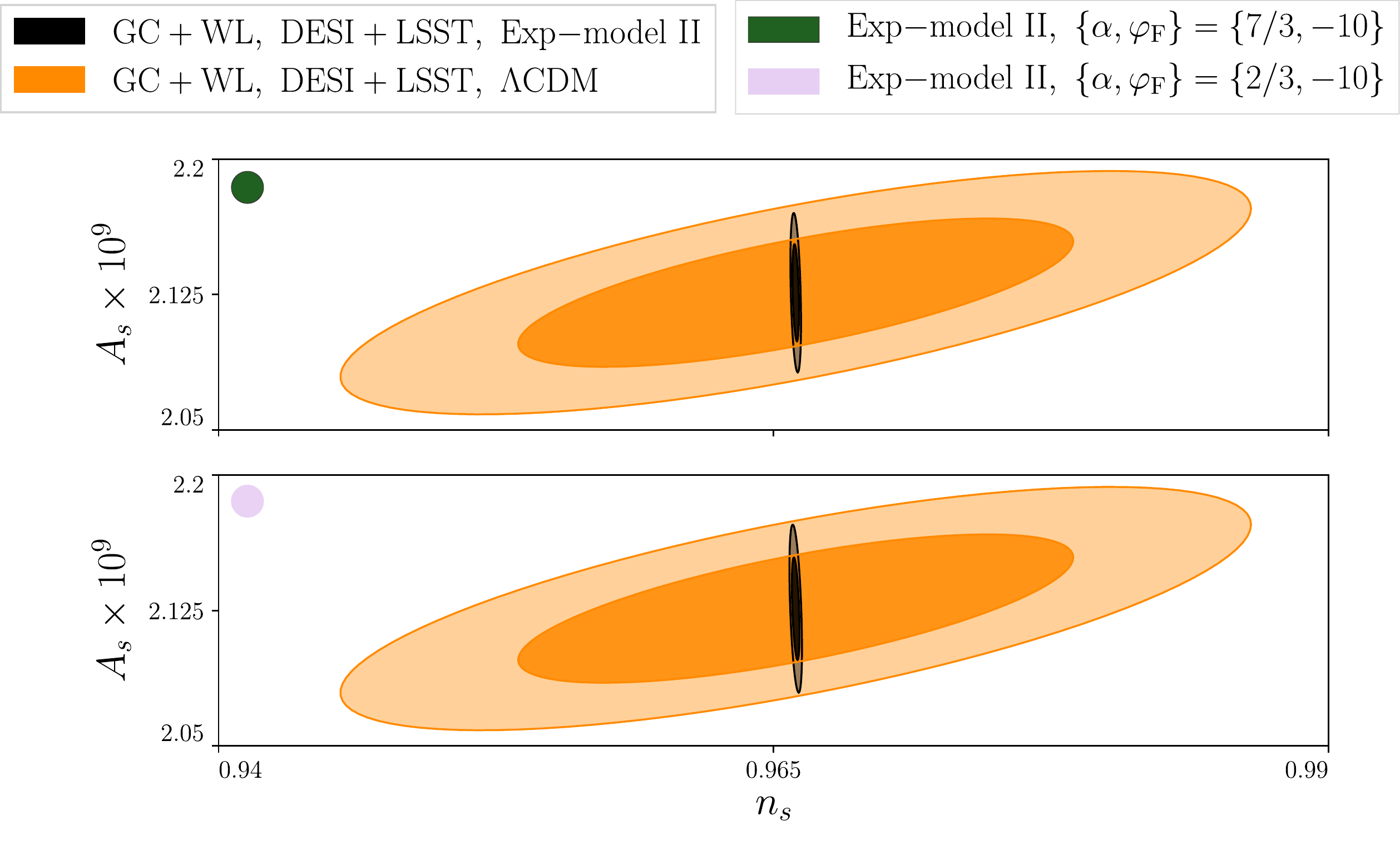}
    \caption{As in Fig.~\ref{fig:DESI-LSST-LCDM-ExpI_As-ns_ellipses}, but for the $\alpha$-attractor quintessential inflation model Exp-model II with $\{\alpha, \varphi_\mathrm{F}\}=\{7/3,-10\}$ (upper panel) and $\{\alpha,\varphi_\mathrm{F}\}=\{2/3,-10\}$ (lower panel).}\label{fig:DESI-LSST-LCDM-ExpII_As-ns_ellipses}
\end{figure*}

The results of our forecast analysis are provided in Fig.~\ref{fig:DESI-LSST-SKA2-ExpII_triplot_ellipses} and Table~\ref{table:constraints}, similarly to the previous case of Exp-model I, where we see that similar conclusions can be made here in terms of the comparison with $\Lambda$CDM, as well as the constraining power of DESI+LSST versus SKA2 and GC versus GC+WL. In particular, the forecasted constraints on the spectral index $n_s$ are extremely tighter in Exp-model II than those assuming $\Lambda$CDM. In order to see this more clearly, in Fig.~\ref{fig:DESI-LSST-LCDM-ExpII_As-ns_ellipses} we show, similarly to the previous cases of Exp-model I, the 2-dimensional Fisher constraints from DESI+LSST on $\mathcal{A}_s$ and $n_s$ for Exp-model II versus $\Lambda$CDM, where again combinations of GC and WL surveys are used.

As another interesting case, let us analyze Exp-model II with $\{\alpha,\varphi_\mathrm{F}\}=\{2/3,-10\}$ for the following reasons. Based on our discussions in Section~\ref{sec:mod-params}, and in particular Figs.~\ref{fig:wDE},~\ref{fig:H} and~\ref{fig:f}, the largest deviation of Exp-model II from $\Lambda$CDM in the entire considered ranges of $\alpha$ and $\varphi_\mathrm{F}$ values corresponds to the case with $\alpha=1/3$ and $\varphi_\mathrm{F}=-10$; note again that here the smallest $\alpha$ provides the largest deviation from $\Lambda$CDM (assuming $\varphi_\mathrm{F}$ is fixed), in contrast to Exp-model I. However, as shown in Ref.~\cite{Akrami:2017cir}, Exp-model II with $\alpha\lesssim0.5$ is already excluded by existing cosmological data. For this reason, we set $\alpha$ to the next smallest value of $2/3$ in the range of values considered in this paper. As for the value of $\varphi_\mathrm{F}$, we keep it the same as in the previous case, i.e., $-10$, in order to maximize deviations of the model's predictions from those of $\Lambda$CDM.

\begin{figure*}
    \centering
    \includegraphics[scale=0.7]{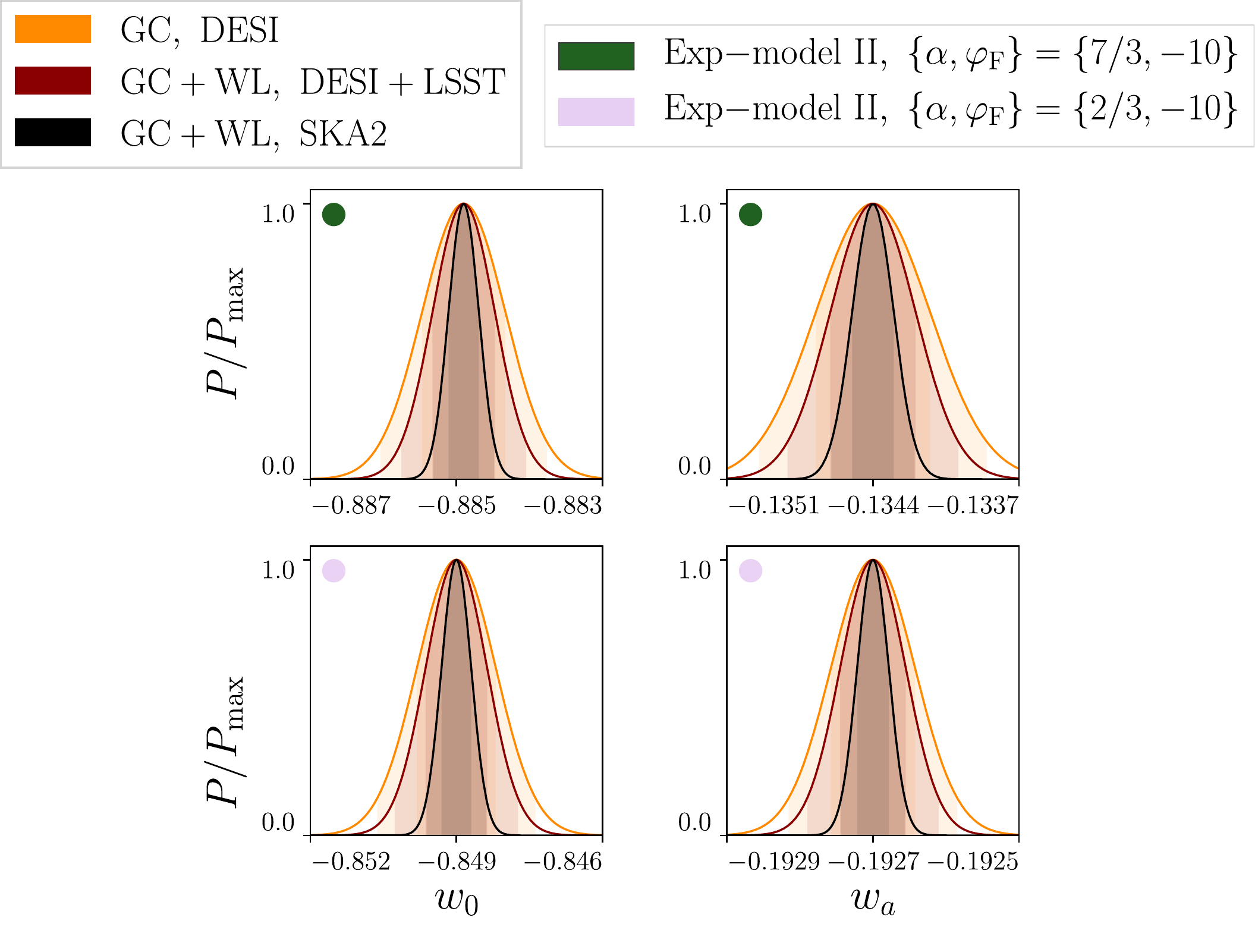}
    \caption{As in Fig.~\ref{fig:DESI-LSST-SKA2-ExpI_1Dplot_w0wa}, but for the $\alpha$-attractor quintessential inflation model Exp-model II with $\{\alpha, \varphi_\mathrm{F}\}=\{7/3,-10\}$ (upper panels) and $\{\alpha,\varphi_\mathrm{F}\}=\{2/3,-10\}$ (lower panels).}\label{fig:DESI-LSST-SKA2-ExpII_1Dplot_w0wa}
\end{figure*}

Our results show that the forecasted constraints on all parameters $\{\Omega_\mathrm{M}^{0}, n_s, \mathcal{A}_s, h\}$ remain very similar to those shown in Fig.~\ref{fig:DESI-LSST-SKA2-ExpII_triplot_ellipses} for the previous case of Exp-model II with $\{\alpha, \varphi_\mathrm{F}\}=\{7/3,-10\}$. We can see this explicitly in Table~\ref{table:constraints}, as well as in Fig.~\ref{fig:DESI-LSST-LCDM-ExpII_As-ns_ellipses} for the Fisher constraints on $\mathcal{A}_s$ and $n_s$ compared to $\Lambda$CDM. However, as far as the dark energy equation of state is concerned, the two cases with $\alpha=7/3$ and $\alpha=2/3$ predict different values for the CPL parameters $w_0$ and $w_a$, as can be seen in Table~\ref{table:fidu-derived} where the fiducial values of $w_0$ and $w_0$ are provided. The forecasted constraints on $w_0$ and $w_a$ for both cases of $\alpha=7/3$ and $\alpha=2/3$ are shown in Fig.~\ref{fig:DESI-LSST-SKA2-ExpII_1Dplot_w0wa} and given in Table~\ref{table:constraints}. The tight constraints on both $w_0$ and $w_a$, in particular the ones expected from SKA2, demonstrate that the two cases of $\alpha=7/3$ and $\alpha=2/3$ are distinguishable from each other and from $\Lambda$CDM with $w_0=-1$ and $w_a=0$.

\begin{figure*}
    \centering
    \includegraphics[scale=0.5]{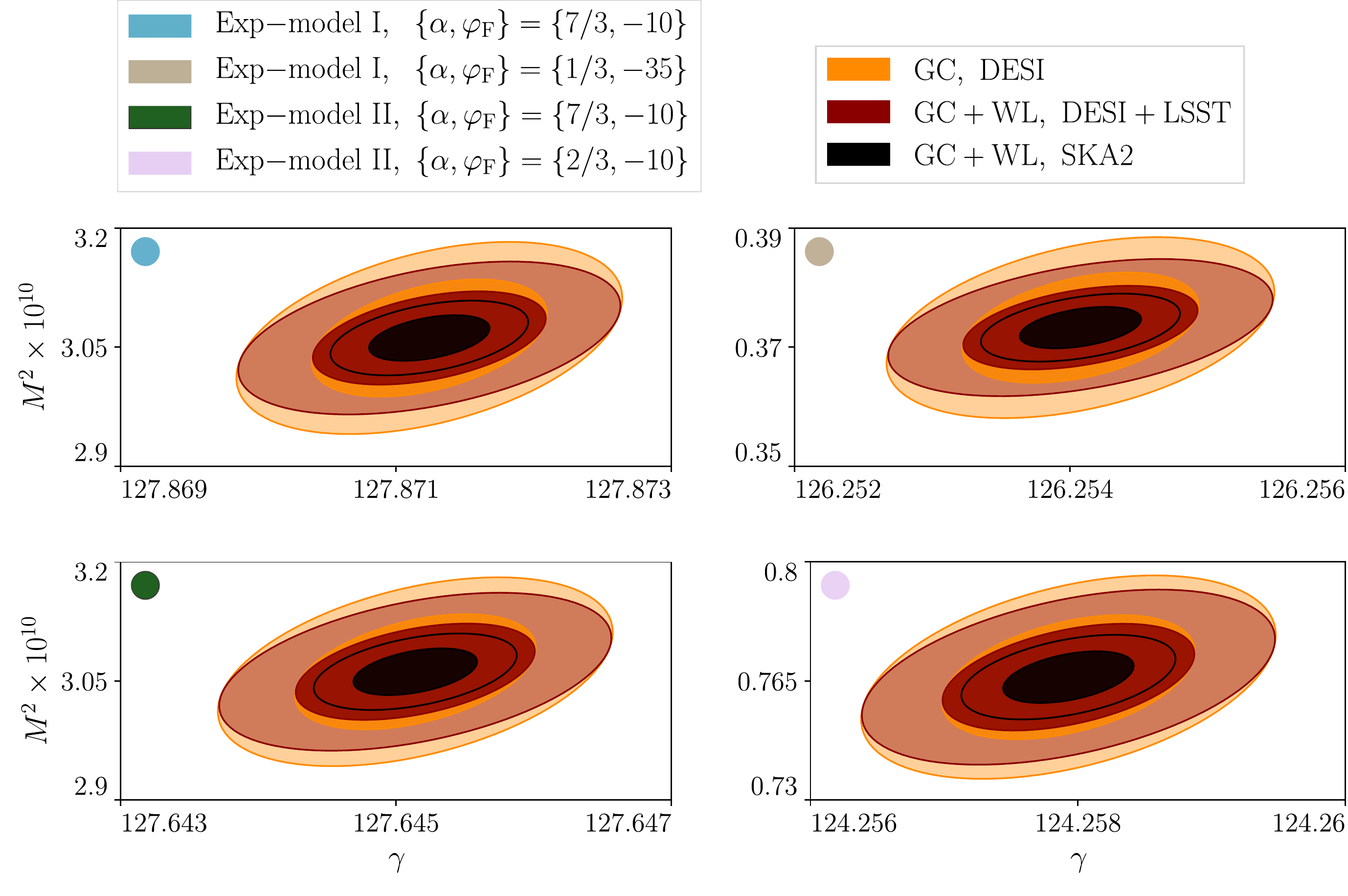}
    \caption{Fisher-matrix $68.3\%$ and $95\%$ confidence constraints on the model parameters $M^2$ and $\gamma$ for $\alpha$-attractor quintessential inflation models Exp-model I with $\{\alpha, \varphi_\mathrm{F}\}=\{7/3,-10\}$ (upper left), Exp-model I with $\{\alpha,\varphi_\mathrm{F}\}=\{1/3,-35\}$ (upper right), Exp-model II with $\{\alpha, \varphi_\mathrm{F}\}=\{7/3,-10\}$ (lower left) and Exp-model II with $\{\alpha,\varphi_\mathrm{F}\}=\{2/3,-10\}$ (lower right). For each panel, we show the results for galaxy clustering (GC) alone, provided by DESI, and the combination of galaxy clustering and weak lensing (GC+WL), provided by either DESI in combination with LSST or SKA2.}\label{fig:M2-gamma}
\end{figure*}

Up to now, we have presented the results of our forecast analysis only in terms of the $\Lambda$CDM parameters $\{\Omega_\mathrm{M}^{0}, n_s, \mathcal{A}_s, h\}$ or the additional parameters $w_0$ and $w_a$ in the CPL extension of $\Lambda$CDM. Even though two of these parameters, $n_s$ and $\mathcal{A}_s$, have played the roles of primary parameters in our Fisher analysis, they do not appear in the theoretical description of $\alpha$-attractors. The specific $\alpha$-attractor models we have studied in this paper are described at the theoretical level by the four model parameters $\alpha$, $\varphi_\mathrm{F}$, $\gamma$ and $M$. We fixed the values of $\alpha$ and $\varphi_\mathrm{F}$ in all our cases, but $\gamma$ and $M$ have been varied. For completeness, we provide in Fig.~\ref{fig:M2-gamma} the obtained forecast constraints on $\gamma$ and $M$ for the four cases discussed above and summarized in Tables~\ref{table:fidu} and~\ref{table:fidu-derived}. Finally, in Table~\ref{table:constraints} we provide $1\sigma$ uncertainties on these two parameters when combinations of GC and WL surveys are used, i.e., for SKA2 and the combination DESI+LSST.

\begin{figure*}
    \centering
    \includegraphics[scale=0.7]{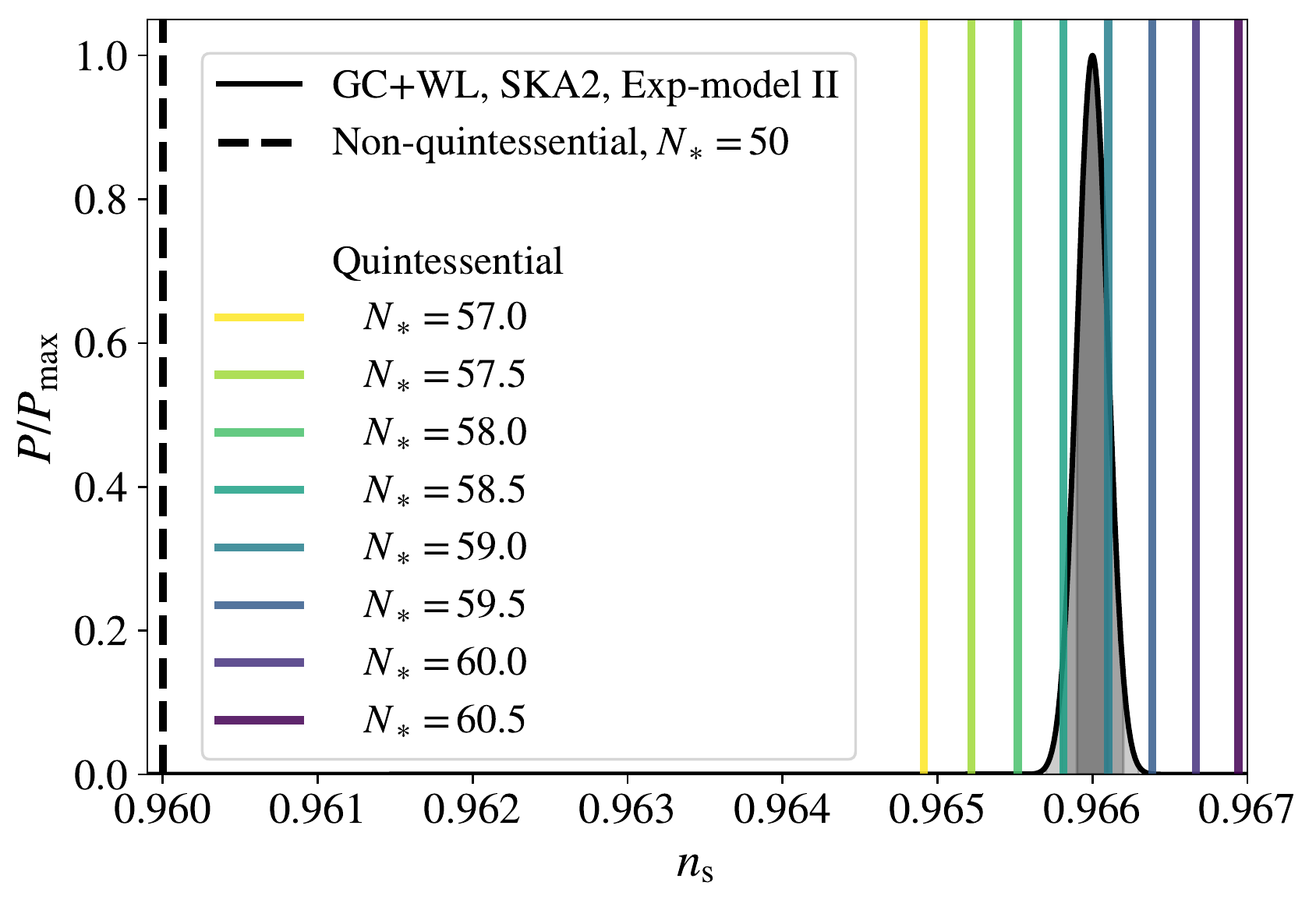}
    \caption{Fisher-matrix 1-dimensional constraint on the spectral index of primordial scalar perturbations $n_s$ for the $\alpha$-attractor quintessential inflation model Exp-model II with $\{\alpha, \varphi_\mathrm{F}\}=\{7/3,-10\}$ and fiducial value of $0.966$ for $n_s$. The forecasted probability density function is the result of the combination of galaxy clustering (GC) and weak lensing (WL) to be measured by SKA2. We have additionally provided theoretical predictions for $n_s$ corresponding to a range of other choices of $N_*$, the number of $e$-foldings between the horizon crossing of modes of interest and the end of inflation, that are allowed by $\alpha$-attractor quintessential inflation, as well as the typical value of $N_*\sim50$ for conventional, non-quintessential models of inflation. The figure demonstrates that future tight constraints on $n_s$ will make it possible to distinguish different models of quintessential inflation from each other and from non-quintessential models of inflation.}\label{fig:ns}
\end{figure*}

Let us now focus on one of the main findings of our forecast analysis, i.e., the extremely tight constraints we expect the next generation of large-scale structure surveys to place on $n_s$ for our $\alpha$-attractor models of quintessential inflation. In Fig.~\ref{fig:ns}, we present the marginalized 1-dimensional constraint on $n_s$ expected from SKA2 (with GC+WL) for Exp-model II with $\{\alpha, \varphi_\mathrm{F}\}=\{7/3,-10\}$ and the fiducial value of $n_s=0.966$. We must first emphasize that this is only a representative case, as the forecasted constraints are almost identically strong for all the $\alpha$-attractor models we have considered in this paper, independently of the type of the model and the values of $\alpha$ and $\varphi_\mathrm{F}$; cf., e.g., Table~\ref{table:constraints}. The fiducial value of $0.966$ for $n_s$ is also a representative value for quintessential $\alpha$-attractor models with typical values of $N_*\sim 60$. Depending on the exact details of the models, in particular properties of the kination phase after the end of inflation and the reheating mechanism, $N_*$ can be slightly lower or higher than $60$~\cite{Dimopoulos:2017zvq,Akrami:2017cir}, while $N_*$ is expected to have values of $\sim50$ for conventional, non-quintessential models of inflation; cf. Section~\ref{sec:background}. We have presented in Fig.~\ref{fig:ns} a few examples of $N_*\sim60$ for quintessential inflation, as well as the typical value of $N_*\sim50$ for non-quintessential inflation. As the figure clearly demonstrates, constraints on $n_s$ expected from upcoming galaxy surveys are so strong that it will soon be possible to know whether a quintessential or non-quintessential mechanism had driven cosmic inflation. It will additionally be possible to obtain detailed information about the exact nature of the inflationary and reheating mechanisms, if it turns out that the Universe underwent a phase of quintessential inflation.

\section{Conclusions}\label{sec:conclusions}

Despite the increasing evidence that the Universe underwent a rapidly accelerating expansion phase at the very early moments of its history, the nature of this period of cosmic inflation is still unknown. Additionally, there is ample undeniable evidence that the current expansion rate of the Universe is increasing, although the nature of this late-time cosmic acceleration is also unknown. A large number of next-generation, Stage IV, cosmological surveys will soon provide us with unprecedentedly precise observations of the formation, evolution and distribution of cosmic large-scale structure, which, in combination with other cosmological probes such as those measuring the cosmic microwave background, are expected to shed light on the mechanisms behind the two periods of accelerating expansion. This will exploit the full potential of the powerful synergistic approach that is already being taken in observational cosmology.

A similar synergistic strategy can be adapted on the theoretical side, i.e., cosmological model building. An intriguing example of this approach is the construction of cosmological models that describe inflation and dark energy in a unified way, through the dynamics of a single degree of freedom and a handful of common model parameters. The class of single-field models of $\alpha$-attractor quintessential inflation is an example of theoretically well-motivated models which have been proven to successfully explain the evolution of the Universe from the earliest to the latest moments. Not only do they set the cosmic initial conditions with distinct predictions for the primordial perturbations, they also link the properties of these initial seeds to the late-time evolution of the Universe and the way dark energy behaves. It is therefore important to know how well these models can be tested by the upcoming cosmological surveys against the standard model of cosmology, where inflation is assumed to be of non-quintessential nature only setting the cosmic initial conditions, as well as the quintessence models of dark energy, where the late-time cosmic acceleration is assumed to be due to a mechanism that is completely separate from the early-Universe physics. One would also like to know how well the parameters of the models will be constrained by the next-generation surveys, and whether or not different models in this class of $\alpha$-attractors can be distinguished from each other.

In this paper, we have provided answers to these questions. After a brief review of the theory of single-field, $\alpha$-attractor, quintessential inflation and the implications of the models for the background evolution of the Universe and the evolution and growth of cosmic structure, we have introduced a set of specific, representative models which capture the main, general properties of $\alpha$-attractor quintessential inflation. For these models, where the potential of the scalar field is assumed to be of an exponential form, we have shown how the number of free parameters can be reduced due to various internal relations between them; this has provided us with a possibility to choose a set of model parameters as close to the parameters of the standard $\Lambda$CDM model as possible.

We have then reviewed the Fisher matrix formalism that we have adapted in this paper for performing our forecast analysis. We have used galaxy clustering and weak gravitational lensing, the two main cosmological probes of the next-generation large-scale structure surveys, to build our forecast machinery, where the observed power spectrum of galaxies and the shear angular power spectrum, as well as their derivatives with respect to the cosmological parameters, have been computed at different fiducial values. We have described our choices of these values and the exact models used in our forecast analysis.

We have considered three major Stage IV galaxy surveys for our studies, the Dark Energy Spectroscopic Instrument (DESI), the Rubin Observatory Legacy Survey of Space and Time (LSST), and the second phase of the Square Kilometre Array (SKA2), and have briefly described their specifications. SKA2, as well as the combination of DESI and LSST, will measure both galaxy clustering and weak lensing, which we have used to estimate the strongest constraints expected from these Stage IV surveys.

Our results have demonstrated that the upcoming large-scale structure surveys considered in this paper will place strong constraints on the parameter space of $\alpha$-attractor quintessential inflation, and will make it possible to test the models against $\Lambda$CDM, where inflation is assumed to be of a non-quintessential nature. We have particularly focused on three important parameters capturing the early- and late-time implications of our models, $n_s$ (the spectral index of primordial curvature perturbations), $w_0$ and $w_a$ (present values of the dark energy equation of state and its time derivative), all of which will be highly constrained by future galaxy surveys.

Our results have shown that, depending on which observables and surveys one considers, $n_s$ is expected to be constrained one or two orders of magnitude more strongly than in $\Lambda$CDM by the same observables and surveys. This tight constraint on $n_s$ is a generic feature of the $\alpha$-attractor models and is independent of their exact details such as the values of model parameters. It provides a powerful way to test the quintessential $\alpha$-attractor models against their non-quintessential versions, as well as gain information about the details of the inflationary and reheating mechanisms, if the models turn out to have been realized in nature. This is because quintessential $\alpha$-attractors predict a value of $n_s\sim0.966$, corresponding to a typical value of $\sim60$ for $N_*$, the number of $e$-foldings between the moment at which modes of interest left the horizon and the end of inflation---this should be compared to a typical value of $N_*\sim50$ for conventional, non-quintessential models.

As for the dark energy equation of state, our results have shown that extremely tight constraints of $\mathcal{O}(10^{-5}\mathrm{-}10^{-4})$ are expected to be placed by Stage IV galaxy surveys on the derived parameters $w_0$ and $w_a$. This should be compared to the expected constraints on these two parameters for the phenomenological CPL (or $w_0w_a$CDM) model, where the standard $\Lambda$CDM model is extended simply by adding the two parameters $w_0$ and $w_a$---the {\it Euclid} space mission~\cite{Laureijs:2011gra}, another Stage IV galaxy survey that we have not considered in this paper, is, for example, expected to measure $w_0$ and $w_a$ within the CPL model with at best the $1\sigma$ uncertainties of $\sigma_{w_0}\approx0.025$ and $\sigma_{w_a}\approx0.092$, respectively~\cite{Blanchard:2019oqi}. The expected strong constraints on the equation of state of dark energy will make it possible, among other things, to constrain the value of $\alpha$, the key parameter of $\alpha$-attractor models, as different values of $\alpha$ give rise to different values of $w_0$ and $w_a$.

Finally, we must emphasize that the forecasted constraints presented in this paper have all been based solely on large-scale structure surveys. Even though, as we have demonstrated, these surveys alone will provide us with invaluable information about the nature of inflation and dark energy, other types of cosmological surveys are also expected to deliver unprecedentedly precise data in the coming years, which will significantly add to our understanding of physics at both early and late times. An obvious example is the upcoming observations of $B$-mode polarization of the cosmic microwave background with the primary objective of measuring the tensor-to-scalar ratio $r$, and therefore, the amount of primordial gravitational waves. Similarly to the scalar spectral index $n_s$, $\alpha$-attractor models provide a universal prediction for $r$,
\begin{equation}
r=3\alpha(1-n_s)^2\,,
\end{equation}
in terms of $\alpha$ and $n_s$. Any positive measurements of $r$ by future CMB experiments will therefore provide us with a measurement of $\alpha$ in the context of $\alpha$-attractor models, given that we will know the value of $n_s$ sufficiently well. Galaxy surveys do not measure $r$, but will provide high-precision measurements of $n_s$, as we have demonstrated in this paper. This means that a combination of the LSS and CMB data will potentially be able to pin down the value of $\alpha$, the central parameter of $\alpha$-attractor models of quintessential inflation.

\acknowledgments Y.A. is supported by LabEx ENS-ICFP: ANR-10-LABX-0010/ANR-10-IDEX-0001-02 PSL*. S.C. is supported by the French Centre National d'Etudes Spatiales (CNES) and the French Centre National de la Recherche Scientifique (CNRS). V.V. is supported by the WPI Research Center Initiative, MEXT, Japan. 

\bibliographystyle{utphys}
\bibliography{refs}

\end{document}